\documentclass[12pt]{article} 
\usepackage{graphicx}
\begin{document}
\baselineskip 10mm

\centerline{\bf Single-qubit operations in the double-donor
structure} \centerline{\bf driven by optical and voltage pulses}
\vskip 2mm

\centerline{Alexander V. Tsukanov}

\vskip 2mm

\centerline{\it Institute of Physics and Technology, Russian
Academy of Sciences} \centerline{\it Nakhimovsky pr. 34, Moscow
117218, Russia} \centerline{\it E-mail: tsukanov@ftian.oivta.ru}

\vskip 4mm

\begin{abstract}
We study theoretically the quantum dynamics of an electron in the
singly-ionized double-donor structure in the semiconductor host
under the influence of laser pulses whose frequencies are close to
structure resonant frequencies. This system can be used as a
charge qubit where the logical states are defined by the lowest
two energy states of the remaining valence electron localized
around one or another donor. The quantum operations are performed
via resonant or Raman-like transitions between the localized
(qubit) states and the excited states delocalized over the
structure, combined with phase shifts between qubit states
generated by voltage pulses. The possibility of realization of
arbitrary single-qubit rotations is demonstrated.
\end{abstract}

\vskip 4mm

PACS number(s): 03.67.Lx, 78.67.-n, 85.35.-p

\vskip 4mm

\newpage

\centerline{\bf I. INTRODUCTION}

  In view of recent progress in the development of the controlled-donor implantation techniques \cite{1},
Kane's paradigm of  the solid-state quantum computations \cite{2}
has gained new insights. The alternative schemes using the orbital
\cite{3,4} or spin \cite{5,6} degrees of freedom of the
donor-bounded electrons to encode the quantum information instead
of the nuclear donor spins, have been proposed. Besides, several
refinements of the original proposal concerning the initialization
\cite{7} and read-out \cite{8,9,10,11} as well as the information
transfer through the quantum networks \cite{12,13,14} have been
made.

In particular, the pair of donors sharing an electron has been
considered as very promising candidate for the solid-state qubit
embodiment \cite{3}. The qubit is presented by the electron
orbital states positioned at the different donors. There are two
main driving mechanisms for the coherent electron evolution
defining a quantum operation on such a qubit. First uses the
electric fields through an application of the adiabatically
switched voltages to the surface gates placed above the donor
structure to modify the confinement potential, thus varying the
electron tunneling rates between neighboring donors
\cite{3,15,16}. The desired state of the qubit is realized by an
appropriate choice of the voltage parameters. The second scheme
relies upon the optical dipole transitions between the
size-quantized one-electron levels induced by the resonant pulses
\cite{17,18}. This technique requires one to irradiate the donors
encapsulated into near-surface semiconductor layers by laser
pulses with frequencies lying in terahertz range. Such kind of
radiation sources is currently under extensive exploration. To all
appearances, the quantum cascade lasers \cite{19} in which photon
generation takes place via electronic intersubband transitions in
semiconductor heterostructures, will be able in near future to
cover this frequency range. As it was shown, an arbitrary
single-qubit operation can be achieved with two simultaneously
switched pulses connecting the qubit states via the excitation of
the state delocalized over the structure. The latter scheme is
likely to be more preferable than the former due to its higher
selectivity and lower field intensity. The successful
implementation of the quantum operations, however, requires one to
provide the high precision in the durations, the frequencies, the
polarizations and the strengths of the pulses together with
reliable control over the delay time and the phase difference
between the pulses. Besides, the use of the intermediate state
lying in the neighborhood of the continuum introduces the
decoherence caused by the ionization and the spontaneous emission
from this state.

Here we propose the way to overcome the difficulties inherent to
the electron charge manipulations by optical means. We show that
the off-resonant laser pulses combined with electrostatic fields,
can be used to generate an arbitrary rotation of the qubit-state
vector as well as to drive an electron between the remote donors.
It is essential that this method is based on the Raman-like
transitions between the localized electron states of an effective
molecular ion where the excited (delocalized over the double-donor
system) states irrespective of their number are used as the
transport channels. As we shall see, the coherent electron
dynamics is described by the simple analytical model. In order to
check analytical results, we have performed numerical simulations
on single-electron dynamics in the effective hydrogen molecular
ion. This system is very popular object for modelling electronic
properties of real donor structures (see, e.g., Refs. \cite{18}
and \cite{20}). The eigenenergies and eigenfunctions of molecular
ion were found from the stationary Schr\"odinger equation and then
used to calculate the set of eigenfrequencies and the matrix
elements of optical dipole transition between eigenstates of
hydrogen molecular ion that define the field-structure interaction
strengths. The time-dependent Schr\"odinger equation generating
the coherent electron evolution in the external fields was
integrated numerically for different values of structure and field
parameters. By comparison of analytical expressions of the Rabi
frequencies for both resonant and off-resonant cases, we found
them to be in good agreement with those extracted from numerical
solutions. This result confirms the validity of approximations
made during the analytical treatment. The pulse and structure
parameters needed for those types of quantum evolution may be
evaluated from the results obtained in this study.

The paper is organized as follows. In Section II we describe
general model of coherent electron dynamics in double-donor
structure under the influence of near-resonant electromagnetic
pulse. The Raman-like off-resonant electronic transitions are
studied analytically in Section III. Besides, in this Section the
possible realization of several basic single-qubit quantum
operations is proposed. The results of the numerical study of
electron dynamics in hydrogen molecular ion are given in Section
IV. The advantages of the off-resonant driving scheme in
comparison with the resonant scheme proposed earlier as well as
some questions concerned with further exploration of
optically-driven donor-based charge qubits, are discussed through
the Section V. We conclude our paper by Section VI.

\centerline{\bf II. MODEL}

We begin with the description of the one-electron double-donor
(DD) structure (Fig. 1 (a)). Let the donors $A$ and $B$ be placed
on the axis $z$ from each other at the distance $R$ (hereafter -
internuclear distance) large enough to consider their ground
orbital states to be isolated. Due to this fact those states may
be used as the qubit states $\left| {0} \right\rangle $ and
$\left| {1} \right\rangle $ (for example, if an electron is
localized on the donor $A (B)$, the qubit is in the state $\left|
{0} \right\rangle $ ($\left| {1} \right\rangle $)). The energy
difference $\Delta  = \varepsilon _1  - \varepsilon _0 $ may be
introduced due to the structural asymmetry caused by the
fabrication process and/or via the bias voltages generated by the
surface gate $V_a$. The coupling between the excited states of
individual donors through the electron tunneling gives rise to the
forming of hybridized states delocalized over the DD structure.
However, if $R\gg a_B^*$ ($a_B^*$ is the effective Bohr radius of
host material) the low-lying excited states are hybridized weakly
and do not participate the two-donor dynamics. We shall be
interested therefore only in the excited states of individual
donors whose wave functions considerably overlap. The
single-electron spectrum of the DD structure is presented by the
sequence of the states $\left\{ {\left| {k} \right\rangle }
\right\}_{exc}$ which for $\Delta=0$ are the doublets composed of
the symmetric and antisymmetric superpositions of isolated donor
states. If $\Delta\neq0$, the spectrum is assumed to be much more
complex, as one can see from Fig. 1 (b). Taking into account the
hydrogen-like spectrum of the isolated donors we expect the
excited states close to the edge of the potential barrier
separating the donors to have the quasi-continuous energy
distribution.

 In what follows we shall study the one-electron
quantum dynamics involving the localized (qubit) states, $\left|
{0} \right\rangle $ and $\left| {1} \right\rangle $, and the
states $\left\{ {\left| {k} \right\rangle } \right\}_{exc}$
delocalized over the structure. Our aim is to choose the field and
structure parameters so that to drive an initial qubit state
$\left| {\Psi \left( t_0 \right)} \right\rangle  = \alpha _0
\left| 0 \right\rangle  + \beta _0 \left| 1 \right\rangle =
\left(\alpha_0, \beta_0 \right)^T$ into the final state $\left|
{\Psi \left( t \right)} \right\rangle  = \alpha \left| 0
\right\rangle + \beta \left| 1 \right\rangle = \left(\alpha, \beta
\right)^T$ with the desired coefficients $\alpha$ and $\beta $.

In the absence of an external field the DD structure is
characterized by the stationary Hamiltonian $H_0$ with the
eigenstates $\left\{ {\left| k \right\rangle } \right\}$ and the
eigenenergies $\left\{ {\varepsilon _k } \right\}$:
\begin{equation}
H_0 \left| k \right\rangle  = \varepsilon _k \left| k
\right\rangle.
\end{equation}
The eigenstates $\left\{ {\left| k \right\rangle } \right\}$ form
the complete orthonormal set so that
\begin{equation}
\sum\limits_k {\left| k \right\rangle \left\langle k \right|}  =
1.
\end{equation}
In the presence of the electromagnetic field the system
Hamiltonian reads
\begin{equation}
H = H_0  - e{\bf{E}}\left( t \right){\bf{r}},
\end{equation}
where $e$ is the electron charge, ${\bf{E}}\left( t \right)$ is
the field strength, $\bf{r}$ is the radius-vector of an electron.
With the help of Eqs. (1) and (2) we rewrite the Eq. (3) in terms
of the projection operators:
\begin{equation}
H = \left( {\sum\limits_k {\left| k \right\rangle \left\langle k
\right|} } \right)H\left( {\sum\limits_m {\left| m \right\rangle
\left\langle m \right|} } \right) = \sum\limits_k {\varepsilon _k
\left| k \right\rangle \left\langle k \right|}  + {\bf{E}}\left( t
\right)\sum\limits_{k,m} {{\bf{d}}_{km} \left| k \right\rangle
\left\langle m \right|} ,
\end{equation}
where ${\bf{d}}_{km}  = \left\langle k \right|-e{\bf{r}}\left| m
\right\rangle $ is the matrix element of optical dipole transition
between the states $\left| {k} \right\rangle $ and $\left| {m}
\right\rangle $. The state vector of the system may be presented
in the form
\begin{equation}
\left| {\Psi \left( t \right)} \right\rangle  = \sum\limits_n {c_n
\left( t \right)e^{ - i\varepsilon _n t} \left| n \right\rangle }
\end{equation}
and is governed by the non-stationary Schr\"odinger equation
\begin{equation}
i\frac{{\partial \left| {\Psi \left( t \right)} \right\rangle
}}{{\partial t}} = H\left| {\Psi \left( t \right)}
 \right\rangle,
\end{equation}
with the initial condition $\left| {\Psi \left( t_0 \right)}
\right\rangle  = \alpha _0 \left| 0 \right\rangle  + \beta _0
\left| 1 \right\rangle $ (hereafter $\hbar  \equiv 1$).

Inserting Eqs. (4) and (5) into Eq. (6), we arrive at the set of
linear differential equations for the probability amplitudes $c_n
\left( t \right)$. We shall only examine the transitions between
the states $\left| {0} \right\rangle $ and $\left| {1}
\right\rangle $ and the states $\left\{ {\left| {k} \right\rangle
} \right\}_{exc}$:
\begin{equation}
 \begin{array}{l}
 i\dot {c_0}  = {\bf{E}}\left( t \right)\sum\limits_k {{\bf{d}}_{0k} c_k e^{ - i\omega _{0k} t} }  \\
 i\dot {c_1}  = {\bf{E}}\left( t \right)\sum\limits_k {{\bf{d}}_{1k} c_k e^{ - i\omega _{1k} t} }  \\
 i\dot {c_k}  = {\bf{E}}\left( t \right)\left( {{\bf{d}}_{0k}^* c_0 e^{i\omega _{0k} t}  + {\bf{d}}_{1k}^* c_1 e^{i\omega _{1k} t} } \right),\,\,k \in \left\{ {k} \right\}_{exc}, \\
 \end{array}
\end{equation}
where $\omega _{0\left( 1 \right)k}  = \varepsilon _k  -
\varepsilon _{0\left( 1 \right)} $.

Let the electromagnetic field imposed on the structure to have (in
the dipole approximation) the form of two phase-locked pulses
\begin{equation}
{\bf{E}}\left( t \right) = {\bf{E}}_0 \left( t \right)\cos \left(
{\omega _0 t + \varphi _0 } \right) + {\bf{E}}_1 \left( t
\right)\cos \left( {\omega _1 t + \varphi _1 } \right),
\end{equation}
where the pulse envelopes ${\bf{E}}_0 \left( t \right) =
{\bf{E}}_0 f_0 \left( t \right),\,\,{\bf{E}}_1 \left( t \right) =
{\bf{E}}_1 f_1 \left( t \right)$ are the slowly-varying (compared
to optical frequencies) time-dependent functions, $\omega _{0,1} $
are the pulse frequencies, and $\varphi _{0,1} $ are the pulse
phases. We require both pulses to be in the two-photon resonance
with the DD structure, i.e., $\varepsilon_0 +  \omega_0 =
\varepsilon_1 +  \omega_1$ or, alternatively,
\begin{equation}
\delta _{0k}  = \delta _{1k}  \equiv \delta _k,
\end{equation}
where $\delta _{0\left( 1 \right)k}  = \omega _{0\left( 1 \right)}
- \omega _{0\left( 1 \right)k} $ is the detuning of the pulse
frequency $\omega _{0\left( 1 \right)} $ from the resonant
frequency $\omega _{0\left( 1 \right)k} $. Note, that we use two
pulses only if $\Delta\ne0$. In the symmetric structure, where
$\Delta=0$, the single pulse ${\bf{E}}\left( t \right)={\bf{E}}_0
\left( t \right)\cos \left( {\omega _0 t + \varphi _0 } \right)$
is enough.

Making use of the rotating-wave approximation (that implies
$\Delta\ll\omega_{0\left( 1 \right)}$ and
$|\delta_k|\ll\omega_{0\left( 1 \right)}$), we obtain from Eq. (7)
the following set:
\begin{equation}
 \begin{array}{l}
 i\dot c_0  = \sum\limits_k {\left[ {\lambda _{0k} \left( t \right) + \mu _{1k} \left( t \right)e^{-i\Delta t} } \right]c_k e^{i\delta _k t} }  \\
 i\dot c_1  = \sum\limits_k {\left[ {\mu _{0k} \left( t \right)e^{ i\Delta t}  + \lambda _{1k} \left( t \right)} \right]c_k e^{i\delta _k t} }  \\
 i\dot c_k  = \left[ {\lambda _{0k}^* \left( t \right) + \mu _{1k}^* \left( t \right)e^{ i\Delta t} } \right]c_0 e^{ - i\delta _k t}  + \\ \,\,\,\,\,\,\,\,\ + \left[ {\mu _{0k}^* \left( t \right)e^{-i\Delta t}  + \lambda _{1k}^* \left( t \right)} \right]c_1 e^{ - i\delta _k t} ,\,\,k \in \left\{ {k} \right\}_{exc}, \\
 \end{array}
\end{equation}
where $\lambda _{0\left( 1 \right)k} \left( t \right) =\lambda
_{0\left( 1 \right)k}f_{0\left( 1 \right)}\left( t
\right)e^{i\varphi _{0\left( 1 \right)} } ,\,\,\mu _{0\left( 1
\right)k} \left( t \right) = \mu _{0\left( 1 \right)k}f_{0\left( 1
\right)}\left( t \right)e^{i\varphi _{0\left( 1 \right)} }$,
$\lambda _{0\left( 1 \right)k}= {{{\bf{d}}_{0\left( 1 \right)k}
{\bf{E}}_{0\left( 1 \right)} } \mathord{\left/
 {\vphantom {{{\bf{d}}_{0\left( 1 \right)k} {\bf{E}}_{0\left( 1 \right)} } 2}} \right.
 \kern-\nulldelimiterspace} 2}$, $\mu _{0\left( 1 \right)k} = {{{\bf{d}}_{1\left( 0 \right)k} {\bf{E}}_{0\left( 1 \right)}  } \mathord{\left/
 {\vphantom {{{\bf{d}}_{1\left( 0 \right)k} {\bf{E}}_{0\left( 1 \right)} } 2}} \right.
 \kern-\nulldelimiterspace} 2}$  and
the identities $\omega _{0\left( 1 \right)}  - \omega _{1\left( 0
\right)k}  = \delta _{1\left( 0 \right)k}  \mp \Delta $ are used.

Eq. (10) describes the dynamical process involving many
three-level excitation schemes that act in parallel. Each of them
is characterized by the set of parameters $\Delta$, $\lambda
_{0\left( 1 \right)k},\,\mu _{0\left( 1 \right)k} $, and
$\delta_k$, where $k \in \left\{ {k} \right\}_{exc}$. We shall
suppose the values of $\lambda_{0(1)k}$ and $\mu_{0(1)k}$ to be of
the same order. Depending on the ratios between these parameters,
$k$-th excitation scheme may be classified in the following way.
First we consider the case of small detunings. If the coupling
coefficients of the optical dipole transitions $\lambda_{0(1)k}$
and the detunings $\delta_k$ satisfy the inequality
\begin{equation}
\left| {\delta _k}\right| ,\Delta  \ll \left| {\lambda _{0k} }
\right|,\left| {\lambda _{1k} } \right|,
\end{equation}
the three-level scheme works in the {\it resonant symmetric}
regime. Instead, the applicability of the {\it resonant
asymmetric} scheme \cite{17} is provided by the condition
\begin{equation}
\left| {\delta _k}\right| \ll \left| {\lambda _{0k} }
\right|,\left| {\lambda _{1k} } \right| \ll \Delta.
\end{equation}
We see that the asymmetry/symmetry of the structure isn't defined
by the presence/absence of the energy difference $\Delta$ only but
by the ratio between $\Delta$ and the coupling coefficients
$\left| {\lambda _{0k} } \right|,\left| {\lambda _{1k} } \right|$
as well. In other words, a driven DD structure can be treated
(relative to the $k$-th transition scheme) as symmetric if the
influence of the parameter $\Delta$, introducing a "static"
asymmetry, is compensated by an appropriate value of the field
strength defined from (11). In this case only one external pulse
is sufficient to excite both transitions \cite{21}.

Next we shall examine the opposite case where the states $\left| 0
\right\rangle$ and $\left| 1 \right\rangle $ are connected through
the off-resonant transitions involving a collection of the excited
states lying at the edge of the potential barrier. Two situations
are possible again:
\begin{equation}
\Delta  \ll \left| {\lambda _{0k} } \right|,\left| {\lambda _{1k}
} \right| \ll \left| {\delta _k}\right|
\end{equation}
 and
\begin{equation}
\left| {\lambda _{0k} } \right|,\left| {\lambda _{1k} } \right|
\ll \left| {\delta _k}\right| ,\Delta.
\end{equation}
The first of these inequalities corresponds to the {\it
off-resonant symmetric} excitation scheme. This situation was
studied in Refs. \cite{22,23} for the double-dot structures with
three and four levels. As it was shown, the set of single-qubit
operations produced by such one-electron dynamics is incomplete,
since in order to realize an arbitrary rotation of the qubit-state
vector, the structure symmetry must be broken. In what follows,
our attention will be focused on both the symmetric case and the
{\it off-resonant asymmetric} case for which the conditions (14)
are satisfied and each pulse drives the transitions between only
one of localized state and the transport states $\left\{ {\left|
{k} \right\rangle } \right\}_{exc}$. This implies also that the
values of $\Delta$ and $\delta_k$ must be rather different from
each other for all $k$ to prevent the single-donor resonant
dynamics.

\centerline{\bf III. THE OFF-RESONANT DYNAMICS}

In this Section we consider two cases of the off-resonant
electronic quantum dynamics in DD structure, i.e., the
off-resonant symmetric dynamics and the off-resonant asymmetric
dynamics. We begin with the second one, initially specifying the
conditions imposed on pulse and structure parameters:
\begin{equation}
\begin{array}{l}
 \left| {\lambda _{0\left( 1 \right)k} } \right|,\,\,\left| {\mu _{0\left( 1 \right)k} } \right| \ll \Delta  \ll \left| {\delta _k } \right|;\,\,\left| {\frac{\partial }{{\partial t}}\lambda _{0\left( 1 \right)k} } \right|,\,\,\left| {\frac{\partial }{{\partial t}}\mu _{0\left( 1 \right)k} } \right| \ll \left| {\lambda _{0\left( 1 \right)k} } \right|^2 ,\,\,\left| {\mu _{0\left( 1 \right)k} } \right|^2 . \\
 \end{array}
\end{equation}
(The case of large $\Delta$ is out of scope of this study; see
Sec. IV D for the reasons).

Strictly speaking, the rotating-wave approximation used in
derivation of Eq. (10) and based on the averaging of Eq. (7) over
time interval $\left[ {\tau  - {\pi  \mathord{\left/
 {\vphantom {\pi  {\omega _0 }}} \right.
 \kern-\nulldelimiterspace} {\omega _0 }},\,\,\tau  + {\pi  \mathord{\left/
 {\vphantom {\pi  {\omega _0 }}} \right.
 \kern-\nulldelimiterspace} {\omega _0 }}} \right]$
 (RWA 1) \cite{24}, cannot be applied to high-lying states for
which the detunings $\delta_k$ become comparable to the pulse
frequencies. Let us define effective "maximal" detuning
$\delta_{\max}\le \omega_0$ that corresponds to the upper bound of
energy range, where RWA 1 is still valid, and average Eq. (10)
over time interval $\left[ {\tau  - {\pi  \mathord{\left/
 {\vphantom {\pi  {\delta _{\max } }}} \right.
 \kern-\nulldelimiterspace} {\delta _{\max } }},\,\,\tau  + {\pi  \mathord{\left/
 {\vphantom {\pi  {\delta _{\max } }}} \right.
 \kern-\nulldelimiterspace} {\delta _{\max } }}} \right]$
 (RWA 2). This step is justified by first of the inequalities (15). After some manipulations, we obtain
two equations for time-averaged functions $c_0$ and $c_1$:
\begin{equation}
\begin{array}{l}
 i\dot c_0  = \sum\limits_k {\frac{1}{{\delta _k }}\left\{ {\left| {\lambda _{0k}  + \mu _{1k} \exp \left( { - i\Delta  \cdot t} \right)} \right|^2 c_0  + \left[ {\lambda _{0k}  + \mu _{1k} \exp \left( { - i\Delta  \cdot t} \right)} \right]\left[ {\lambda _{1k}  + \mu _{0k} \exp \left( {i\Delta  \cdot t} \right)} \right]^* c_1 } \right\}} , \\
 i\dot c_1  = \sum\limits_k {\frac{1}{{\delta _k }}\left\{ {\left| {\lambda _{1k}  + \mu _{0k} \exp \left( {i\Delta  \cdot t} \right)} \right|^2 c_1  + \left[ {\lambda _{0k}  + \mu _{1k} \exp \left( { - i\Delta  \cdot t} \right)} \right]^* \left[ {\lambda _{1k}  + \mu _{0k} \exp \left( {i\Delta  \cdot t} \right)} \right]c_0 } \right\}} . \\
 \end{array}
\end{equation}
Again, careful consideration of averaging procedure requires us to
handle only with the detunings that are sufficiently small in
comparison with $\delta_{\max}$. It is easy to see that the
equations (16) do not contain the fast-oscillating terms
$c_{k\ne0,1}$. Therefore, these equations describe effective
two-state dynamics involving only the localized states $\left|0
\right\rangle$ and $\left|1 \right\rangle$. From third equation of
the set (10) we conclude that the functions $c_{k\ne0,1}$ have the
amplitudes $\sim|\lambda_{0k}/\delta_k|<<1$ and oscillate at the
frequencies $\sim|\delta_k|$. The reduction of set like Eq. (10)
to the two equations for slowly-varying probability amplitudes
$c_0$ and $c_1$ is known in atomic optics as the adiabatic
elimination procedure \cite{25} - \cite{27}. To proceed further,
let us average the equations (16) over the time interval $\left[
{\tau  - {\pi \mathord{\left/
 {\vphantom {\pi  \Delta }} \right.
 \kern-\nulldelimiterspace} \Delta },\tau  + {\pi  \mathord{\left/
 {\vphantom {\pi  \Delta }} \right.
 \kern-\nulldelimiterspace} \Delta }} \right]$
 (RWA 3) on which all of the time-dependent
functions except $\exp \left( { \pm i\Delta t} \right)$ may be
replaced by their mean values so that after integration we arrive
at the following set of equations:
\begin{equation}
\begin{array}{l}
 i\dot{c_0}  = \sum\limits_k {\frac{1}{{\delta _k }}\left\{ {\left[ {\left| {\lambda _{0k} \left( t \right)} \right|^2  + \left| {\mu _{1k} \left( t \right)} \right|^2 } \right]c_0  + \lambda _{0k} \left( t \right)\lambda _{1k}^* \left( t \right)c_1 } \right\}} , \\
 i\dot{c_1}  = \sum\limits_k {\frac{1}{{\delta _k }}\left\{ {\lambda _{0k}^* \left( t \right)\lambda _{1k}^{} \left( t \right)c_0  + \left[ {\left| {\lambda _{1k} \left( t \right)} \right|^2  + \left| {\mu _{0k} \left( t \right)} \right|^2 } \right]c_1 } \right\}} . \\
 \end{array}
 \end{equation}

Now we return to the symmetric driving scheme that is simpler than
the asymmetric one. By setting $\Delta=0$ and ${\bf{E}}_1=0$, we
transform the set of Eq. (10) into
\begin{equation}
\begin{array}{l}
 i\dot {c_0}  = \sum\limits_k {\lambda _{0k} \left( t \right) b_k }  \\
 i\dot {c_1}  = \sum\limits_k {\lambda _{1k} \left( t \right) b_k }  \\
 i\dot {b_k} = - \delta _k b_k  + \lambda _{0k}^* \left( t \right)c_0  + \lambda _{1k}^* \left( t \right)c_1 ,\,\,k \in \left\{ {k} \right\}_{exc}, \\
 \end{array}
\end{equation}
where $b_k  = c_k \exp \left( {  i\delta _k t} \right),\,k \in
\left\{ {k} \right\}_{exc} $ and we redefine the coupling
coefficients as $\lambda _{0(1)k} \left( t \right) = \lambda
_{0(1)k} f_0\left( t \right) e^{i\varphi _0}$, $\lambda _{0(1)k} =
{{{\bf{d}}_{0(1)k} {\bf{E}}_0 } \mathord{\left/
 {\vphantom {{{\bf{d}}_{0(1)k} {\bf{E}}_0 } 2}} \right.
 \kern-\nulldelimiterspace} 2}$
. The inequalities (13) allow one to apply the adiabatic
elimination procedure to the intermediate levels $\left\{ {\left|
{k} \right\rangle } \right\}_{exc}$:
\begin{equation}
\dot {b_k } \approx 0,\,\,b_k  \approx {{\left[ {\lambda _{0k}^*
\left( t \right)c_0  + \lambda _{1k}^* \left( t \right)c_1 }
\right]} \mathord{\left/
 {\vphantom {{\left[ {\lambda _{0k}^* \left( t \right)c_0  + \lambda _{1k}^* \left( t \right)c_1 } \right]} {\delta _k }}} \right.
 \kern-\nulldelimiterspace} {\delta _k }},\,\,k \in \left\{ {k}
 \right\}_{exc}.
\end{equation}

The equations for two remaining probability amplitudes $c_0$ and
$c_1$  in the matrix form for both asymmetric and symmetric
dynamics read
\begin{equation}
i\frac{\partial }{{\partial t}}\left( {\begin{array}{*{20}c}
   {c_0 }  \\
   {c_1 }  \\
\end{array}} \right) = \left( {\begin{array}{*{20}c}
   {\Lambda _{0} \left( t \right)} & {\Lambda _{2} \left( t \right)}  \\
   {\Lambda _{2}^* \left( t \right)} & {\Lambda _{1} \left( t \right)}  \\
\end{array}} \right)\left( {\begin{array}{*{20}c}
   {c_0 }  \\
   {c_1 }  \\
\end{array}} \right),
\end{equation}
where for symmetric case one has $\Lambda _{0} \left( t \right) =
\Lambda _{0} f_0^2 \left( t \right)$, $\Lambda _{1} \left( t
\right) = \Lambda _{1} f_0^2 \left( t \right)$, $\Lambda _{2}
\left( t \right) = \Lambda _{2}
 f_0^2 \left( t
\right)$ and
\begin{equation}
\Lambda _{0}  = \sum\limits_k {{{\left| {\lambda _{0k} } \right|^2
} \mathord{\left/
 {\vphantom {{\left| {\lambda _{0k} } \right|^2 } {\delta _k }}} \right.
 \kern-\nulldelimiterspace} {\delta _k }}} ,
\Lambda _{1}  = \sum\limits_k {{{\left| {\lambda _{1k} } \right|^2
} \mathord{\left/
 {\vphantom {{\left| {\lambda _{1k} } \right|^2 } {\delta _k }}} \right.
 \kern-\nulldelimiterspace} {\delta _k }}},
\Lambda _{2}  = \sum\limits_k {{{\lambda _{0k} \lambda _{1k}^* }
\mathord{\left/
 {\vphantom {{\lambda _{0k} \lambda _{1k}^* } {\delta _k }}} \right.
 \kern-\nulldelimiterspace} {\delta _k }}},
\end{equation}
 whereas for
 asymmetric case the functions entering into Eq. (20) are given by
 expressions
 $\Lambda _0 \left( t \right) = \sum\limits_k {{{\left[ {\left|
{\lambda _{0k} \left( t \right)} \right|^2  + \left| {\mu _{1k}
\left( t \right)} \right|^2 } \right]} \mathord{\left/
 {\vphantom {{\left[ {\left| {\lambda _{0k} \left( t \right)} \right|^2  + \left| {\mu _{1k} \left( t \right)} \right|^2 } \right]} {\delta _k }}} \right.
 \kern-\nulldelimiterspace} {\delta _k }}} $,
  $\Lambda _1 \left( t \right) = \sum\limits_k {{{\left[ {\left|
{\lambda _{1k} \left( t \right)} \right|^2  + \left| {\mu _{0k}
\left( t \right)} \right|^2 } \right]} \mathord{\left/
 {\vphantom {{\left[ {\left| {\lambda _{1k} \left( t \right)} \right|^2  + \left| {\mu _{0k} \left( t \right)} \right|^2 } \right]} {\delta _k }}} \right.
 \kern-\nulldelimiterspace} {\delta _k }}} $, and $\Lambda _{2}\left( t \right) = \sum\limits_k {{{\lambda _{0k}\left( t \right) \lambda _{1k}^*\left( t \right) }
\mathord{\left/
 {\vphantom {{\lambda _{0k}\left( t \right) \lambda _{1k}^* \left( t \right)} {\delta _k }}} \right.
 \kern-\nulldelimiterspace} {\delta _k }}} $. It allows us to treat both cases in the same manner,
 however, keeping in mind that for symmetric
structure $\Lambda_0\left( t \right)=\Lambda_1\left( t \right)$,
since $\lambda_{0k}=\pm\lambda_{1k}$ for the state $\left|k
\right\rangle$ whose electronic wave function possesses even/odd
parity relative to the center of DD structure, that does not take
place for asymmetric structure. Below we present the general
solution of Eq. (20) that will describe both types of off-resonant
electron dynamics.

The eigenstates and the eigenenergies of the matrix in right-hand
side of Eq. (20) may be written as
\begin{equation}
\left\{ \begin{array}{l}
 \left|  +  \right\rangle  = e^{i\arg \left[ {\Lambda _{2} \left( t \right)} \right]} \cos \left[ {{{\Theta \left( t \right)} \mathord{\left/
 {\vphantom {{\Theta \left( t \right)} 2}} \right.
 \kern-\nulldelimiterspace} 2}} \right]\left| 0 \right\rangle  + \sin \left[ {{{\Theta \left( t \right)} \mathord{\left/
 {\vphantom {{\Theta \left( t \right)} 2}} \right.
 \kern-\nulldelimiterspace} 2}} \right]\left| 1 \right\rangle  \\
 \left|  -  \right\rangle  = e^{i\arg \left[ {\Lambda _{2} \left( t \right)} \right]} \sin \left[ {{{\Theta \left( t \right)} \mathord{\left/
 {\vphantom {{\Theta \left( t \right)} 2}} \right.
 \kern-\nulldelimiterspace} 2}} \right]\left| 0 \right\rangle  - \cos \left[ {{{\Theta \left( t \right)} \mathord{\left/
 {\vphantom {{\Theta \left( t \right)} 2}} \right.
 \kern-\nulldelimiterspace} 2}} \right]\left| 1 \right\rangle  \\
 \end{array} \right.
\end{equation}
and
\begin{equation}
E_ \pm  \left( t \right) = {{\left[ {\Lambda _{0} \left( t \right)
+ \Lambda _{1} \left( t \right)} \right]} \mathord{\left/
 {\vphantom {{\left[ {\Lambda _{0} \left( t \right) + \Lambda _{1} \left( t \right)} \right]} 2}} \right.
 \kern-\nulldelimiterspace} 2} \pm \Omega \left( t \right),
\end{equation}
respectively, where
\begin{equation}
\cos \left[ {\Theta \left( t \right)} \right] = {{\left[ {\Lambda
_{0} \left( t \right) - \Lambda _{1} \left( t \right)} \right]}
\mathord{\left/
 {\vphantom {{\left[ {\Lambda _{0} \left( t \right) - \Lambda _{1} \left( t \right)} \right]} {2\Omega \left( t \right)}}} \right.
 \kern-\nulldelimiterspace} {2\Omega \left( t \right)}},\,\,\,\,\sin \left[ {\Theta \left( t \right)} \right] = {{\left| {\Lambda _{2} \left( t \right)} \right|} \mathord{\left/
 {\vphantom {{\left| {\Lambda _{2} \left( t \right)} \right|} {\Omega \left( t \right)}}} \right.
 \kern-\nulldelimiterspace} {\Omega \left( t \right)}},
\end{equation}
and
\begin{equation}
\Omega \left( t \right) = \sqrt {{{\left[ {\Lambda _{0} \left( t
\right) - \Lambda _{1} \left( t \right)} \right]^2 }
\mathord{\left/
 {\vphantom {{\left[ {\Lambda _{0} \left( t \right) - \Lambda _{1} \left( t \right)} \right]^2 } 4}} \right.
 \kern-\nulldelimiterspace} 4} + \left| {\Lambda _{2} \left( t \right)} \right|^2 }
\end{equation}
is the instantaneous Rabi frequency. Using the unitary
transformation
\begin{equation}
D\left( t \right) = \left( {\begin{array}{*{20}c}
   {e^{i\arg \left[ {\Lambda _{2} \left( t \right)} \right]} \cos \left[ {{{\Theta \left( t \right)} \mathord{\left/
 {\vphantom {{\Theta \left( t \right)} 2}} \right.
 \kern-\nulldelimiterspace} 2}} \right]} & {e^{i\arg \left[ {\Lambda _{2} \left( t \right)} \right]} \sin \left[ {{{\Theta \left( t \right)} \mathord{\left/
 {\vphantom {{\Theta \left( t \right)} 2}} \right.
 \kern-\nulldelimiterspace} 2}} \right]}  \\
   {\sin \left[ {{{\Theta \left( t \right)} \mathord{\left/
 {\vphantom {{\Theta \left( t \right)} 2}} \right.
 \kern-\nulldelimiterspace} 2}} \right]} & { - \cos \left[ {{{\Theta \left( t \right)} \mathord{\left/
 {\vphantom {{\Theta \left( t \right)} 2}} \right.
 \kern-\nulldelimiterspace} 2}} \right]}  \\
\end{array}} \right)
\end{equation}
we represent the state vector in the instantaneous basis $\left\{
{\left|  +  \right\rangle ,\left|  -  \right\rangle } \right\}$ as
\begin{equation}
\left| \Phi \left( t \right)  \right\rangle  = a_ + \left( t
\right) \left|  + \right\rangle  + a_ - \left( t \right) \left|  -
\right\rangle ,\,\,\,\left| \Psi \left( t \right) \right\rangle =
D\left( t \right)\left| \Phi \left( t \right) \right\rangle
\end{equation}
and rewrite Eq. (20) in the new basis as
\begin{equation}
i\frac{\partial }{{\partial t}}\left( {\begin{array}{*{20}c}
   {a_ +  }  \\
   {a_ -  }  \\
\end{array}} \right) = \left( {\begin{array}{*{20}c}
   {E_ +  } & {{{\dot \Theta } \mathord{\left/
 {\vphantom {{\dot \Theta } 2}} \right.
 \kern-\nulldelimiterspace} 2}}  \\
   { - {{\dot \Theta } \mathord{\left/
 {\vphantom {{\dot \Theta } 2}} \right.
 \kern-\nulldelimiterspace} 2}} & {E_ -  }  \\
\end{array}} \right)\left( {\begin{array}{*{20}c}
   {a_ +  }  \\
   {a_ -  }  \\
\end{array}} \right),
\end{equation}
where
\begin{equation}
\dot \Theta \left( t \right) = \frac{{\left| {\left[ {\dot \Lambda
_0 \left( t \right) - \dot \Lambda _1 \left( t \right)}
\right]\left| {\Lambda _2 \left( t \right)} \right| - \left| {\dot
\Lambda _{2} \left( t \right)} \right|\left[ {\Lambda _0 \left( t
\right) - \Lambda _1 \left( t \right)} \right]} \right|}}{{\left[
{\Lambda _0 \left( t \right) - \Lambda _1 \left( t \right)}
\right]^2 /4  + \left| {\Lambda _2 \left( t \right)} \right|^2 }}.
\end{equation}
Here we restrict our interest by the diagonal evolution followed
from the choice of the system parameters for which $\dot \Theta
\left( t \right) \ll E_{\pm} \left( t \right)$. In this case the
solution of Eq. (28) is straightforward:
\begin{equation}
\begin{array}{l}
 a_ +  \left( t \right) = a_ +  \left( {t_0 } \right)\exp \left[ { - i\int\limits_{t_0 }^t {E_ +  \left( {t'} \right)dt'} } \right], \\
 a_ -  \left( t \right) = a_ -  \left( {t_0 } \right)\exp \left[ { - i\int\limits_{t_0 }^t {E_ -  \left( {t'} \right)dt'} } \right]. \\
 \end{array}
\end{equation}
With the help of equations (26), (27), and (30) we may write down
the expression for the evolution matrix of the qubit-state vector
$\left| {\Psi \left( t \right)} \right\rangle $ in the laboratory
frame:
\begin{equation}
\begin{array}{l}
 \left| {\Psi \left( t \right)} \right\rangle  = U\left| {\Psi \left( t_0 \right)} \right\rangle , \\
 U = \left( {\begin{array}{*{20}c}
   {e^{ - i\varepsilon _0 t} } & 0  \\
   0 & {e^{ - i\varepsilon _1 t} }  \\
\end{array}} \right)D\left( t \right)\left( {\begin{array}{*{20}c}
   {e^{ - i\int\limits_{t_0 }^t {E_ +  \left( {t'} \right)dt'} } } & 0  \\
   0 & {e^{ - i\int\limits_{t_0 }^t {E_ -  \left( {t'} \right)dt'} } }  \\
\end{array}} \right)D^\dag  \left( t_0 \right) =  \\
 \,\,\,\,\,\, = e^{ - i\left[ {\varepsilon _0 t + \varphi _\Lambda  \left( t \right)} \right]} \left( {\begin{array}{*{20}c}
   {u_{00} } & {u_{01} }  \\
   {u_{10} e^{ - i\Delta t} } & {u_{11} e^{ - i\Delta t} }  \\
\end{array}} \right), \\
 \end{array}
\end{equation}
where
\begin{equation}
\begin{array}{l}
 u_{00}  = u_{11}^*  = e^{ - i\tilde \Omega \left( t \right)} \cos \left[ {{{\Theta \left( t \right)} \mathord{\left/
 {\vphantom {{\Theta \left( t \right)} 2}} \right.
 \kern-\nulldelimiterspace} 2}} \right]\cos \left[ {{{\Theta \left( {t_0 } \right)} \mathord{\left/
 {\vphantom {{\Theta \left( {t_0 } \right)} 2}} \right.
 \kern-\nulldelimiterspace} 2}} \right] + e^{i\tilde \Omega \left( t \right)} \sin \left[ {{{\Theta \left( t \right)} \mathord{\left/
 {\vphantom {{\Theta \left( t \right)} 2}} \right.
 \kern-\nulldelimiterspace} 2}} \right]\sin \left[ {{{\Theta \left( {t_0 } \right)} \mathord{\left/
 {\vphantom {{\Theta \left( {t_0 } \right)} 2}} \right.
 \kern-\nulldelimiterspace} 2}} \right], \\
 u_{01}  =  - u_{10}^*  = \\
 =e^{i\arg \left[ {\Lambda _{2} \left( t \right)} \right]} \left\{ {e^{ - i\tilde \Omega \left( t \right)} \cos \left[ {{{\Theta \left( t \right)} \mathord{\left/
 {\vphantom {{\Theta \left( t \right)} 2}} \right.
 \kern-\nulldelimiterspace} 2}} \right]\sin \left[ {{{\Theta \left( {t_0 } \right)} \mathord{\left/
 {\vphantom {{\Theta \left( {t_0 } \right)} 2}} \right.
 \kern-\nulldelimiterspace} 2}} \right] - e^{i\tilde \Omega \left( t \right)} \sin \left[ {{{\Theta \left( t \right)} \mathord{\left/
 {\vphantom {{\Theta \left( t \right)} 2}} \right.
 \kern-\nulldelimiterspace} 2}} \right]\cos \left[ {{{\Theta \left( {t_0 } \right)} \mathord{\left/
 {\vphantom {{\Theta \left( {t_0 } \right)} 2}} \right.
 \kern-\nulldelimiterspace} 2}} \right]} \right\}, \\
 \end{array}
\end{equation}
and
\begin{equation}
\tilde \Omega \left( t \right) = \int\limits_{t_0 }^t {\Omega
\left( {t'} \right)dt'} ,\,\,\,\varphi _\Lambda  \left( t \right)
= \int\limits_{t_0 }^t {{{\left[ {\Lambda _{0} \left( {t'} \right)
+ \Lambda _{1} \left( {t'} \right)} \right]} \mathord{\left/
 {\vphantom {{\left[ {\Lambda _{0} \left( {t'} \right) + \Lambda _{1} \left( {t'} \right)} \right]} 2}} \right.
 \kern-\nulldelimiterspace} 2}dt'} .
\end{equation}
The expressions (31) - (33) describe the effective two-level
dynamics that corresponds to the continuous evolution of the qubit
state vector on the Bloch sphere. In the remainder of this section
we show how to choose the pulse and structure parameters in order
to realize the most important single-qubit gates. We illustrate
the qubit state engineering by considering a particular case of
the driving pulses sharing the same time dependence, i.e., $f_0
\left( t \right) = f_1 \left( t \right) \equiv f\left( t \right)$.
The condition $\dot \Theta  = 0$ is then satisfied and the
components of the evolution matrix (31) take the form
\begin{equation}
\begin{array}{l}
 u_{00}  = u_{11}^*  = \cos \left[ {\tilde \Omega \left( t \right)} \right] - i\cos \left( {\Theta _0 } \right)\sin \left[ {\tilde \Omega \left( t \right)} \right], \\
 u_{01}  =  - u_{10}^*  =  - ie^{i\arg \left( {\Lambda _{2} } \right)} \sin \left( {\Theta _0 } \right)\sin \left[ {\tilde \Omega \left( t \right)} \right], \\
 \end{array}
\end{equation}
where $\Theta _0  = \arcsin \left[ {{{\left| {\Lambda _{2} }
\right| } \mathord{\left/
 {\vphantom {{\left| {\Lambda _{2} } \right| } {\sqrt {{{\left( {\Lambda _{0}  - \Lambda _{1} } \right)^2 } \mathord{\left/
 {\vphantom {{\left( {\Lambda _{00}  - \Lambda _{11} } \right)^2 } 4}} \right.
 \kern-\nulldelimiterspace} 4} + \left| {\Lambda _{2} } \right|^2 } }}} \right.
 \kern-\nulldelimiterspace} {\sqrt {{{\left( {\Lambda _{0}  - \Lambda _{1} } \right)^2 } \mathord{\left/
 {\vphantom {{\left( {\Lambda _{0}  - \Lambda _{1} } \right)^2 } 4}} \right.
 \kern-\nulldelimiterspace} 4} + \left| {\Lambda _{2} } \right|^2 } }}} \right]$.
The dynamics described by the equations (34) is sufficient to
generate an arbitrary single-qubit rotation on the Bloch sphere.
For example, the quantum operations such as NOT ($\sigma _ x$):
$\left(\alpha_0, \beta_0 \right)^T\rightarrow \left(\beta_0,
\alpha_0 \right)^T$; PHASE ($\sigma _ z$): $\left(\alpha_0,
\beta_0 \right)^T\rightarrow \left(\alpha_0, -\beta_0 \right)^T$;
and Hadamard ($H$): $\left(\alpha_0, \beta_0 \right)^T\rightarrow
\left[ \left( \alpha_0 + \beta_0 \right)/ \sqrt{2}, \left(
\alpha_0 - \beta_0 \right)/ \sqrt{2}  \right]^T$ can be realized
(up to the common phase) given the following choices of the pulse
- structure parameters:
\begin{equation}
\tilde \Omega \left( T \right) = {\pi  \mathord{\left/
 {\vphantom {\pi  {2 + \pi k,\,\,\,}}} \right.
 \kern-\nulldelimiterspace} {2 + \pi k,\,\,\,}}T\Delta  = 2\pi l,\,\,\,\arg \left( {\Lambda _{2} } \right) = 2\pi m
\end{equation}
and
\begin{equation}
\Theta_0 \left( {\sigma _X } \right) = {\pi  \mathord{\left/
 {\vphantom {\pi  2}} \right.
 \kern-\nulldelimiterspace} 2} + 2\pi n,\,\,\,\Theta_0 \left( {\sigma _Z } \right) = \pi n,\,\,\,\Theta_0 \left( H \right) = {\pi  \mathord{\left/
 {\vphantom {\pi  4}} \right.
 \kern-\nulldelimiterspace} 4} + \pi n,
\end{equation}
respectively. Here $k,l,m,$ and $n$ are the integers and $T$ is
the pulse duration. Of course, this is not a unique parameter
choice to attain the above quantum operations. Careful estimation
of the operation times requires the detailed knowledge of the
energy spectrum of DD structure and the values of
$\lambda_{0(1)k}$.

\centerline{\bf IV. A HYDROGEN MOLECULAR ION: NUMERICAL STUDY }

We visualize our results with the help of model, where the
energies and wave functions of an electron, bounded in the DD
structure, are approximated by the eigenenergies and
eigenfunctions of effective hydrogen molecular ion H$^+_2$, for
which the effective Bohr radius $a^*_B=\epsilon(m_0/m^*)a_B$ and
the effective Rydberg energy $Ry^*=[m^*/(\epsilon^2m_0)]Ry$
contain the information about real solid-state environment via the
electronic effective mass $m^*$ and the dielectric constant
$\epsilon$ ($Ry=13.6057$ eV is Rydberg energy,
$a_B=0.5292\times10^{-10}$ m is the Bohr radius, and
$m_0=9.1094\times10^{-31}$ kg is the free electron mass). In
practice, the single-valley approximation was applied to study the
phonon-induced decoherence of electron in Si:P$^+_2$ structure
\cite{20} and the effect of surface gate on single phosphorous
donor in the silicon \cite{28}, as well as to calculate the Rabi
frequency of resonant electron transfer between localized states
in Si:P$^+_2$ structure in three-level approximation \cite{18}.
The comparison of the results of calculations, where intervalley
interference is taken into account, with those neglecting such
effects, demonstrates qualitative agreement between them
(especially in presence of an external field, see \cite{16}). In
our study, we are mostly interested in validation of the
approximations, made during the paper to reveal the specific
features of resonant and off-resonant electron dynamics in a
multilevel structure. We expect, that the dynamical analysis
developed above and confirmed numerically below for effective
single-valley-approximated structure, can be applied to real
solid-state DD structures as well.

In what follows we show how to find the eigenenergies and
eigenfunctions of hydrogen molecular ion (HMI) without LCAO
approximation. We calculate the matrix elements of optical dipole
transition and study their behavior under axial electric field. We
shall work with atomic units (a.u.), keeping for energy 1 a.u. =
2$Ry^*=E_D$ ($E_D$ is the effective ionization energy) and for
distance 1 a.u. = $a^*_B$.

\centerline{\bf A. Eigenenergies and eigenfunctions of HMI at zero
bias field}

There exist several ways how to compute the eigenenergies and
eigenfunctions of HMI. We consider one of them using variable
separation in the time-independent Schr\"odinger equation followed
by representation of solution in new variables via appropriate
series expansions (see the work \cite{29} and references therein).
It is well known that the Schr\"odinger equation for HMI
\begin{equation}
\left( {\frac{1}{2}\nabla ^2  + E + \frac{1}{{r_A }} +
\frac{1}{{r_B }}} \right)\Psi \left( {\bf{r}} \right) = 0,
\end{equation}
where $r_{A(B)}$ denotes the distance between an electron and atom
$A(B)$, is separable in prolate spheroidal coordinates $\xi =
{{\left( {r_A + r_B } \right)} \mathord{\left/
 {\vphantom {{\left( {r_A  + r_B } \right)} R}} \right.
 \kern-\nulldelimiterspace} R},\,\,\eta  = {{\left( {r_A  - r_B } \right)} \mathord{\left/
 {\vphantom {{\left( {r_A  - r_B } \right)} R}} \right.
 \kern-\nulldelimiterspace} R},\,\,\varphi  = \arctan \left( {{y \mathord{\left/
 {\vphantom {y x}} \right.
 \kern-\nulldelimiterspace} x}} \right)$ ($\,1 \le \xi  \le \infty
 $, $ - 1 \le \eta  \le 1$,  $0 \le \varphi  \le 2\pi $ )
 and a (non-normalized) stationary electronic wave
function of HMI can be written in the form $\Psi \left( {\xi ,\eta
,\varphi } \right) = \Xi \left( \xi  \right){\rm H}\left( \eta
\right)\Phi \left( \varphi \right)$, where functions $\Xi \left(
\xi  \right)$ and ${\rm H}\left( \eta  \right)$ meet the
generalized radial and angular spheroidal wave equations,
respectively, and the azimuthal function $\Phi \left( \varphi
\right)$ is the same as for hydrogen atom: $\Phi \left( \varphi
\right) = {{\exp \left( {im\varphi } \right)} \mathord{\left/
 {\vphantom {{\exp \left( {im\varphi } \right)} {\sqrt {2\pi } }}} \right.
 \kern-\nulldelimiterspace} {\sqrt {2\pi } }}$, where $m = 0, \pm 1, \pm
 2,..$ is the azimuthal (or magnetic) quantum number. The most useful
expansion for ${\rm H}\left( \eta  \right)$ is in a series of
associated Legendre polynomials, whereas for $\Xi \left( \xi
\right)$ the power series expansion is applied. Substituting these
functions into corresponding wave equations and requiring them to
be minimal solutions of these equations, we obtain the expansion
coefficients and separation constants $C_{\xi}$ and $C_{\eta}$
that at fixed $m$ depend continuously on the energy parameter $E$
and on the internuclear distance $R$. To obtain electronic
eigenenergies we now have to find the set of separation constants
common for both radial and angular solutions for given $m$ and
$R$. Graphically, these can be sought as intersection points of
curves from sets $\left\{C_{\xi}(E)\right\}$ and
$\left\{C_{\eta}(E)\right\}$ and corresponding values of parameter
$E$ will be the eigenenergies of HMI (for more details, see the
Ref. \cite{29}).

The HMI eigenstates is completely specified by the set of quantum
numbers $\left( {N_\xi  ,N_\eta  ,m} \right)$, where $N_\xi
(N_\eta)$ equals to the number of zeros in radial (angular)
function $\Xi \left( \xi  \right)$ (${\rm H}\left( \eta \right)$).
On the other hand, in the united atom limit ${R \to 0}$ the
quantum numbers $\left( {n_u,l,m} \right)$ are used, where $n_u$
specifies the energy level, and $l$ the angular momentum. Both
sets are related by the formulas $N_\xi   = n_u - l - 1,\,\,N_\eta
= l - m$. In what follows we shall use the second one for the
state classification as more instructive, i.e., $\Psi \left( {\xi
,\eta ,\varphi } \right)\leftrightarrow \left| n_ulm\right\rangle
$. According to standard atomic notation, the letters $s$, $p$,
$d$, $f$, $g$, $h$, ... are used to denote the values of $l=0, 1,
2, 3, ...$ in united atom limiting case, and the greek letters
$\sigma$, $\pi$, $\delta$, $\phi$, ... denote the values of $m$.
We can also classify the states of homonuclear HMI by the parity
under the transformation ${\bf{r}} \to  - {\bf{r}}$, namely, the
state with symmetrical (antisymmetrical) wave function will be
supplied with subscript $g$ ($u$).

Making use of computational framework sketched above, we have
found the eigenenergies and eigenfunctions of discrete part of HMI
energy spectrum. At Fig. 2, the eigenenergies of first 20
$\sigma$-states ($m=0$) are plotted as functions of internuclear
distance $R$ for $20<R<40$ (on the whole, we found the
eigenenergies of 64 states with $m=0$). We also computed the
eigenenergies of states from subspaces with $m\ne0$, however, only
$\sigma$-states are relevant for the dynamical description of
electron evolution in the electric fields polarized along $z$ axis
that will be in the focus of further consideration. The HMI
eigenenergies are found with absolute accuracy 10$^{-15}$ a.u.
that is enough for this study, but the described algorithm enables
one to calculate them with absolute accuracy 10$^{-20}$ a.u. or
even higher \cite{29}. From Fig. 2 (a) it is seen that two
low-lying states of HMI, i.e., $\left| {1s\sigma _g }
\right\rangle $ and $\left| {2p\sigma _u } \right\rangle $, are
almost degenerate, and the energy difference $\Delta _{1s\sigma _g
- 2p\sigma _u }  = \varepsilon \left( {2p\sigma _u } \right) -
\varepsilon \left( {1s\sigma _g } \right)$ decreases exponentially
with $R$. The dependence of the electronic tunneling time $\tau
_{1s\sigma _g  - 2p\sigma _u }  = {\hbar \mathord{\left/
 {\vphantom {\hbar  {\Delta _{1s\sigma _g  - 2p\sigma _u } }}} \right.
 \kern-\nulldelimiterspace} {\Delta _{1s\sigma _g  - 2p\sigma _u } }}$
(here $\hbar=6.582\times10^{-16}$ eV$\times$ s) between these
states on $R$ can be seen at Fig. 3, where the parameters
$E_D$=-45 meV and $a^*_B$=1.22 nm, taken from Ref. \cite{16} for
Si:P$^+_2$ in single valley approximation, are used. From this
Figure we obtain an estimation on $\tau _{1s\sigma _g  - 2p\sigma
_u } $ for $R\approx$30 nm ($\approx$20 a.u.) to be of order of 1
$\mu$s that is enough for carrying out some proof-of principle
experiments, whereas for $R\approx$50 nm ($\approx$40 a.u.) one
has for the tunneling time $\tau _{1s\sigma _g - 2p\sigma _u
}\approx 10^2$ s that would enable to run quantum algorithms. The
states $\left| {1s\sigma _g } \right\rangle $ and $\left|
{2p\sigma _u } \right\rangle $ are presented by even and odd
superpositions of states localized on donors $A$ and $B$.
Inversely, the logical (localized) states can be expressed by even
and odd superpositions of the states $\left| {1s\sigma _g }
\right\rangle $ and $\left| {2p\sigma _u } \right\rangle $ of HMI:
$\left| {0\left( 1 \right)} \right\rangle = \left| {1s^{A\left( B
\right)} } \right\rangle  = {{\left( {\left| {1s\sigma _g }
\right\rangle  \pm \left| {2p\sigma _u } \right\rangle } \right)}
\mathord{\left/
 {\vphantom {{\left( {\left| {1s\sigma _g } \right\rangle  \pm \left| {2p\sigma _u } \right\rangle } \right)} {\sqrt 2 }}} \right.
 \kern-\nulldelimiterspace} {\sqrt 2 }}$.

The collection of low-lying electronic excited states falls into
the subbands characterized at $R>>$1 by the principal quantum
number $n$ of isolated hydrogen atom. The sets of HMI eigenstates,
whose energies are pictured at Fig. 2 (b), namely, $\left\{
{\left| {4f\sigma _u } \right\rangle ,\left| {3d\sigma _g }
\right\rangle , \left| {3p\sigma _u } \right\rangle ,\left|
{2s\sigma _g } \right\rangle } \right\}$, $\left\{ {\left|
{6h\sigma _u } \right\rangle ,\left| {5g\sigma _g } \right\rangle
, \left| {5f\sigma _u } \right\rangle ,\left| {4d\sigma _g }
\right\rangle \left| {4p\sigma _u } \right\rangle , \left|
{3s\sigma _g } \right\rangle } \right\}$, and $\left\{ {\left|
{7j\sigma _u } \right\rangle ,\left| {6h\sigma _g } \right\rangle
,..} \right\}$ (not labelled), correspond to $n=$2, $n=$3, and
$n=$4, respectively. In their turn, these subbands can be
subdivided further into the doublets $\left| {ks\sigma _g }
\right\rangle  - \left| {\left( {k + 1} \right)p\sigma _u }
\right\rangle $, $k$=2,3,..; $\left| {kd\sigma _g } \right\rangle
- \left| {\left( {k + 1} \right)f\sigma _u } \right\rangle $,
$k$=3,4,..; and so on. The degree of hybridization/localization
can be evaluated from state quantum numbers. For example, the
pairs of states with $l=s,p$ become degenerate at comparatively
small internuclear distances. The doublets of states with $l=d,f$
are well resolved up to $R\approx$30 and collapse quickly for
larger $R$. We observe almost complete dissociation of HMI states,
pertaining to the subband with $n=$2, into two pairs of states
$\left| {2s^{A(B)} } \right\rangle  = {{\left( {\left| {2s\sigma
_g } \right\rangle \pm \left| {3p\sigma _u } \right\rangle  +
\left| {3d\sigma _g } \right\rangle  \pm \left| {4f\sigma _u }
\right\rangle } \right)} \mathord{\left/
 {\vphantom {{\left( {\left| {2s\sigma _g } \right\rangle  \pm \left| {3p\sigma _u } \right\rangle  + \left| {3d\sigma _g } \right\rangle  \pm \left| {4f\sigma _u } \right\rangle } \right)} 2}} \right.
 \kern-\nulldelimiterspace} 2}$
and $\left| {2p\sigma ^{A(B)} } \right\rangle  = {{\left( {\left|
{2s\sigma _g } \right\rangle  \pm \left| {3p\sigma _u }
\right\rangle  - \left| {3d\sigma _g } \right\rangle  \mp \left|
{4f\sigma _u } \right\rangle } \right)} \mathord{\left/
 {\vphantom {{\left( {\left| {2s\sigma _g } \right\rangle  \pm \left| {3p\sigma _u } \right\rangle  - \left| {3d\sigma _g } \right\rangle  \mp \left| {4f\sigma _u } \right\rangle } \right)} 2}} \right.
 \kern-\nulldelimiterspace} 2}$ of isolated hydrogen atoms.
Taking the average over the energies of states pertaining to the
subbbands with $n=1$ and $n=2$ and adding the energy of
internuclear repulsion $\varepsilon_{nucl}=1/R$, we arrive at
isolated hydrogen atom energies $\varepsilon(n)=1/2n^2$. On the
other hand, the states with higher values of $l$ remain
non-degenerate on the whole interval of $R$ under consideration.
One can therefore treat those states as reliable transport
channels to drive an electron between logical states. As we shall
see below, the states $\left|{5g\sigma_g} \right\rangle$ and
$\left| {6h\sigma_u}\right\rangle$ (whose energies are plotted as
thick red lines at Fig. 2 (b)) meet the conditions imposed on the
choice of transport states by the high excitation selectivity and
appropriate dipole coupling strength requirements. The states,
whose energies are plotted as thick blue lines, may be exploited
as transport ones either at $R<20$ ($\left|{3d\sigma_g}
\right\rangle$ and $\left| {4f\sigma_u}\right\rangle$) or at
$R>30$ ($\left|{6g\sigma_g} \right\rangle$, $\left|
{7h\sigma_u}\right\rangle$), $\left|{7i\sigma_g} \right\rangle$).

\centerline{\bf B. Matrix elements of optical dipole transition at
zero bias field}

The numerical solution of Eq. (6) implies the knowledge of matrix
elements $\left\{ {\bf{d}}_{km} \right\}$ of optical dipole
transition (ODT) between all states entering into the Eq. (4). In
our simulation we restrict ourselves by consideration of electric
fields polarized along $z$ axis, so that the field-structure
interaction Hamiltonian in Eq. (3) reads $- eE \left( t \right)
z$. In this case, the matrix elements of ODT  between states
$\left| k\right\rangle$ and $\left| m\right\rangle$ can be
calculated in spheroidal coordinates through the expression
\begin{equation}
d_{km}  = \frac{1}{{\sqrt {N_k N_m } }}\left\{ {\int\limits_{ -
1}^1 {d\eta } \int\limits_1^\infty  {d\xi \,\left[ {{\rm H}_k
\left( \eta  \right)\Xi _k \left( \xi  \right)} \right]\left( { -
{{eR \xi \eta} \mathord{\left/
 {\vphantom {{eR \xi \eta} 2}} \right.
 \kern-\nulldelimiterspace} 2}} \right)\left[ {{\rm H}_m \left( \eta  \right)\Xi _m \left( \xi  \right)} \right]J\left( {\xi, \eta } \right)} } \right\},
\end{equation}
\begin{equation}
N_{k\left( m \right)}  = \int\limits_{ - 1}^1 {d\eta }
\int\limits_1^\infty  {d\xi \,\left[ {{\rm H}_{k\left( m \right)}
\left( \eta  \right)\Xi _{k\left( m \right)} \left( \xi  \right)}
\right]^2 J\left( {\xi, \eta } \right)} ,\,\,\,\,\,J\left( {\xi,
\eta } \right) = \xi ^2  - \eta ^2,
\end{equation}
where $J\left( {\xi, \eta } \right)$ is the Jacobian of the
transformation from cartesian frame to spheroidal frame, and the
relation $z = \left( {{R \mathord{\left/
 {\vphantom {R 2}} \right.
 \kern-\nulldelimiterspace} 2}} \right)\xi \eta $ is used.
(The wave functions ${\rm H}_{k\left( m \right)} \left( \eta
\right)$ and $\Xi _{k\left( m \right)}$ of discrete part of HMI
energy spectrum are real functions, hence the ODT matrix elements
will be real as well). According to the selection rule $\Delta m =
0$, imposed by the structure axial symmetry and by the choice of
the field polarization, and keeping in mind that logical states
pertain to the $\sigma$-subspace of HMI eigenstates, only the
transitions among the states with $m=0$ are relevant. Besides, the
analysis of Eq. (38) supplies us with other selection rule
standing for allowed transitions $\Delta l$ to be an odd number.
In other words, the states from subspaces $\left\{
{s,\,d,\,g,\,i,\,..} \right\}$ with even $l$ are dipole-coupled to
the states from subspaces $\left\{ {p,\,f,\,h,\,j,\,..} \right\}$
with odd $l$. It means that ODT selection rule upon $l$, that
follows from spatial symmetry of HMI, turns out to be relaxed in
comparison with that of hydrogen-like atom for which $\Delta l=\pm
1$.

Using the equations (38) and (39), we have calculated the ODT
matrix elements between all pairs of 64 low-lying $\sigma$-states
of HMI as functions of internuclear distance $R$. In what follows,
they will be used to define the right-hand side of Eq. (6). At the
same time, in Eq. (21) we proceed with matrix elements between
logical states $\left| 0 \right\rangle $ and $\left| 1
\right\rangle $ and excited ones. These matrix elements can be
derived from expression $d_{0\left( 1 \right)\,k} = {{\left(
{d_{1s\sigma _g \,k}  \pm d_{2p\sigma _u \,k} } \right)}
\mathord{\left/
 {\vphantom {{\left( {d_{1s\sigma _g \,k}  \pm d_{2p\sigma _u \,k} } \right)} {\sqrt 2 }}} \right.
 \kern-\nulldelimiterspace} {\sqrt 2 }}$. The dependencies of ${d_{0k} }$ (1 a.u. = $ea_B^*$)
on $R$ are given at Figs. 4 (a) - (c) for subbands with $n=$2, 3,
4. Note that in symmetric HMI $d_{0k} = d_{1k} $, if $k$ denotes a
state with even parity, and $d_{0k}  = -d_{1k} $, if $k$ denotes a
state with odd parity. The calculation of matrix elements between
localized states $\left| {1s^{A,B} } \right\rangle $ from subband
with $n=$1 (logical subspace) and localized states $\left|
{2s^{A,B} } \right\rangle $ and $\left| {2p\sigma ^{A,B} }
\right\rangle $ from subband with $n=$2, yields at $R=38$ $d_{1s -
2s}^{A,B} \approx 0$ and $d_{1s - 2p\sigma }^{A,B} \approx 0.746$
that coincide with values $d_{z\,1s - 2s }=0$ and $d_{z\,1s -
2p\sigma } = \frac{{2^8 }}{{3^5 \sqrt 2 }}\left( {ea_B^*} \right)
 \approx 0.745$ for hydrogen atom.

\centerline{\bf C. Resonant and off-resonant electron dynamics in
zero-bias case}

Here we present the results of numerical simulations on coherent
electron evolution, paying attention to performing $\sigma_X$
qubit-state vector rotation. For simplicity, we work with square
pulses and $f(t)$ entering into Eq. (34) is the step function. To
drive an electronic population in symmetric structure such as HMI,
a single pulse is only needed, thus we integrate Eq. (6) with
Hamiltonian $H = H_0 - eE_0 z\cos \left( {\omega _0 t} \right)$,
and initial conditions corresponding to the localization of an
electron into logical state $\left| 0 \right\rangle $, are
$c_{1s\sigma _g } \left( 0 \right) = c_{2p\sigma _u } \left( 0
\right) = {1 \mathord{\left/
 {\vphantom {1 {\sqrt 2 }}} \right.
 \kern-\nulldelimiterspace} {\sqrt 2 }}$. Our goal is to calculate the
frequencies of Rabi oscillations between logical states and to
estimate the degree of population leakage from logical subspace
for different values of internuclear distance $R$, pulse strength
$E_0$, and pulse frequency $\omega_0$. For this purpose, we
analyze the probabilities $p_0\left( T \right)$ and $p_1\left( T
\right)$ ($p_{0\left( 1 \right)} \left( T \right) = {{\left|
{c_{1s\sigma _g } \left( T \right) \pm c_{2p\sigma _u } \left( T
\right)} \right|^2 } \mathord{\left/
 {\vphantom {{\left| {c_{1s\sigma _g } \left( T \right) \pm c_{2p\sigma _u } \left( T \right)} \right|^2 } 2}} \right.
 \kern-\nulldelimiterspace} 2}$) to find electron into the logical states $\left| 0
\right\rangle $ and $\left| 1 \right\rangle $, together with total
probability $p_{tr} \left( T \right) = \sum\limits_{k \ne 0,1}
{p_k \left( T \right)} $ of electron to be out of logical
subspace, versus the pulse duration $T$.

Let us introduce the dimensionless field energy $\varepsilon
_{field}  = {{eE_0 a_B^* } \mathord{\left/
 {\vphantom {{eE_0 a_B^* } {2Ry^* }}} \right.
 \kern-\nulldelimiterspace} {2Ry^* }}$ and the dimensionless pulse duration
$T = {{2Ry^* t} \mathord{\left/
 {\vphantom {{2Ry^* t} \hbar }} \right.
 \kern-\nulldelimiterspace} \hbar }$.
 According to the analysis given in Sec. II, the coupling coefficients that in atomic units
take the form $\lambda _{0\left( 1 \right)k}  = {{\varepsilon
_{field} d_{0\left( 1 \right)k} } \mathord{\left/
 {\vphantom {{\varepsilon _{field} d_{0\left( 1 \right)k} } 2}} \right.
 \kern-\nulldelimiterspace} 2}$, have to satisfy the requirements imposed by a concrete
optical excitation regime. If one applies a resonant driving
scheme, where HMI state $\left| r \right\rangle $ is used as
transport state, the detuning of pulse frequency $\omega _0$ from
resonant frequency $\omega _{0r} $ must be much smaller than the
value of coupling coefficient $|\lambda _{0 r}|$. In its turn,
$|\lambda _{0 r}|$ must be much smaller than $\omega _{rr+1}$ and
$\omega _{r-1r}$ in order to minimize population leakage into the
states nearest-in-energy to transport state. Instead, the
exploitation of strongly-detuned pulses (Raman scheme) implies the
values of coupling coefficients $|\lambda _{0 k}|$ for all excited
states to be much smaller than corresponding detunings. It is easy
to see that one could attain this condition taking pulse strength
as small as possible and detunings as large as possible. However,
this brings about considerable reduction in Rabi frequency. In so
far, we shall be interested in determination of optimal pulse
parameter set, that would amount to rapid and robust
implementation of quantum operations.

We begin with resonant driving scheme that, being realized in
symmetric structure, may be considered as auxiliary one, since it
is only able to inverse the population of logical states at
discrete set of pulse durations when the electron is concentrated
into logical subspace. The electronic resonant population transfer
in three-level and in four-level structures was studied
theoretically in Refs. \cite{21} and \cite{23}, where the
probabilities $p_0$, $p_1$, and $p_{tr}$ were found at exact
resonance (zero detuning from transport level) to be
\begin{equation}
 p_0 \left( T \right) = \cos ^4 \left( {\frac{{E_0 d_{0tr} T}}{4}}
\right),\,p_1 \left( T \right) = \sin ^4 \left( {\frac{{E_0
d_{0tr} T}}{4}} \right),\,p_{tr} \left( T \right) =
\frac{1}{2}\sin ^2 \left( {\frac{{E_0 d_{0tr} T}}{2}} \right).
\end{equation}
One can observe from Fig. 5 that these expressions are in
excellent agreement with numerical curves. We have performed our
simulations on resonant dynamics for several values of
$\varepsilon_{field}$ (different pulse strengths $E_0$) and pulse
frequencies $\omega_0$, matching resonant frequencies of HMI
(different $d_{0tr}$) in order to define the dependency of Rabi
frequency on these parameters. We have revealed that the pulses
with $\varepsilon_{field}<0.003$ being tuned on resonance with
states $\left| 5g\sigma_g \right\rangle $ or $\left| 6h\sigma_u
\right\rangle $, provide good selectivity, low population leakage
at the end of NOT operation ($p_1 \left( {T_{NOT}^{} = {\pi
\mathord{\left/
 {\vphantom {\pi  {2\Omega _R }}} \right.
 \kern-\nulldelimiterspace} {2\Omega _R }}} \right) > 0.999$
), and quite high speed of electronic transfer
($T_{NOT}\approx10^4$ that is of order of hundreds of picoseconds
for Si:P$_2^+$ structure).We have also performed numerical
simulations on the electron dynamics for $R=22$ in order to check
the possibility of using the states from HMI subband with $n=2$ as
transport ones. As it was expected, the resonant population
transfer between logical states via excitation of one of the
states $\left\{\left|3d\sigma_g\right\rangle, \left|4f\sigma_u
\right\rangle\right\}$ do really take place. However, large dipole
moments for those transitions (see Fig. 4 (a)) require the field
energy to be sufficiently low since the energy spacings between
subband levels are still small. As a result, the time needed for
complete electron transfer is of the same order as it was for
larger $R$, where the states from third and fourth subbands,
characterized by smaller values of ODT matrix elements but
demonstrating higher resolution over energy, play role of
transport channels. Thus the formula $\Omega _R  = {{\left| {E_0
d_{0tr} } \right|} \mathord{\left/
 {\vphantom {{\left| {E_0 d_{0tr} } \right|} 4}} \right.
 \kern-\nulldelimiterspace} 4} = {{\left| {\lambda _{0tr} } \right|} \mathord{\left/
 {\vphantom {{\left| {\lambda _{0tr} } \right|} 2}} \right.
 \kern-\nulldelimiterspace} 2}$ approximates with high accuracy
the Rabi frequency for resonant electronic transfer for $20<R<40$.

Now we present the results of calculations for the off-resonant
scheme. Typical curves for $p_0$, $p_1$, and $p_{tr}$, reflecting
coherent electron dynamics driven by strongly detuned pulses, are
plotted at Fig. 6. They demonstrate essentially two-level
oscillatory behavior where the population is mainly localized in
logical subspace. This picture differs from that obtained for
resonant scenario by sharp decrease in the amplitude of population
$p_{tr}$ combined with considerable increase in the oscillation
period of $p_0$ and $p_1$ (by order of magnitude or more).
According to Eqs. (21), (31), (33) and (34), in the symmetric
structure $\Lambda_0-\Lambda_1=0$, $\Theta_0=\pi/2$, and
\begin{equation}
p_0 \left( T \right) \approx \cos ^2 \left( {\Omega _R T}
\right),\,\,p_1 \left( T \right) \approx \sin ^2 \left( {\Omega _R
T} \right),
\end{equation}
where $\Omega _R  = \left| {\Lambda _2 } \right|$ (see Eq. (25)).
Relative to the transport states, we are only able to estimate
under adiabatic elimination an order of magnitude of total
probability as $p_{tr}  \sim \left( {{{\lambda _{0k} }
\mathord{\left/
 {\vphantom {{\lambda _{0k} } {\delta _{0k} }}} \right.
 \kern-\nulldelimiterspace} {\delta _{0k} }}} \right)_{\max }^2 <<1$
and its oscillation frequency $\Omega _{tr}  \sim \left( {{{\delta
_{0k}^2 } \mathord{\left/
 {\vphantom {{\delta _{0k}^2 } {\lambda _{0k} }}} \right.
 \kern-\nulldelimiterspace} {\lambda _{0k} }}} \right)_{\max } >>\Omega _R$,
so that $p_{tr}$ exhibits fast oscillations with small amplitude.

In order to compare the Rabi frequencies found numerically with
those calculated within analytical framework, we plot both data
types on Fig. 7 versus pulse frequency (shifted by
$\varepsilon_0$) for $R=30$ and $R=38$ and for the field energies
$\varepsilon_{field}$=0.003 and $\varepsilon_{field}$=0.005. To
extract numerical values of $\left| {\Lambda _2 } \right|$, we fit
the curves for $p_1 \left( T \right)$ by squared sine function and
then define $\Omega _R$ as the sine frequency. These values, drawn
as filled circles, correlate well with analytical results of Eq.
(21) pictured by solid curves. Since the approximation applied in
derivation of Eqs. (20) and (21) do not allow one to work in near
neighborhood of HMI energy levels, we have left empty the
intervals around the levels marking them by vertical dotted lines.
We shall regard the oscillations as two-level ones if the depth of
modulations of $p_0$ and $p_1$, arising from the non-resonant
population of excited states, is smaller than 0.01. At Fig. 7,
there are several points (enclosed by open red squares) satisfying
to this conventional criterion. Other numerical data points
correspond to the $p_{tr}$ maxima  (or modulation depths) ranging
from 0.01 to 0.05. For the points located at the boundaries of the
intervals, on which approximate solution is valid, the maximum
values of $p_{tr}$ turn out to be 0.05 or higher. With further
approaching of the pulse frequency to one of HMI resonant
frequencies, the off-resonant oscillatory picture transforms into
resonant one. The off-resonant dynamics at $R=22$ (not shown) is
quite similar to that presented at Fig. 7. Again, almost ideal
two-level oscillations are obtained if we tune the pulse frequency
into the middle of doublet
$\left\{\left|5g\sigma_g\right\rangle,\left|6h\sigma_u
\right\rangle\right\}$. With that, the use of doublet
$\left\{\left|3d\sigma_g\right\rangle, \left|4f\sigma_u
\right\rangle\right\}$ becomes possible provided that the pulse
strength is not high ($\varepsilon_{field}\le 0.001$). As it will
be shown below, we cannot efficiently operate with such pulses
since the Rabi frequency of qubit rotations appears to be very
small.

Note that the character of oscillations of total population of
excited states is regular enough in both resonant and off-resonant
cases; this observation can be explained by assumption that only
several excited states participate the dynamics. Such an
explanation is obvious for resonant excitation scheme, but in
off-resonant case, where all excited states are equivalent in
dynamical sense, rigorous arguments are needed. To find them, let
us consider the sum $\Lambda_2$ in more details. As we have
mentioned before, the excited states of HMI can be classified by
tunnel coupling strength between symmetric and antisymmetric
states pertaining to the same doublet. For weakly hybridized
states $\left|k_1 \right\rangle $ and $\left|k_2 \right\rangle $,
provided that the value of tunnel splitting is much smaller than
pulse detuning from one of these states, we can write
$\delta_{k_1}\approx\delta_{k_2}$. We suppose, that the pulse
energy bandwidth is much larger than $|\delta_{k_1}-\delta_{k_2}|$
so we are disable to resolve the doublet states under
consideration. Furthermore, in this case
$\lambda_{0k_1}\approx\lambda_{0k_2}$,
$\lambda_{0k_1}\approx\lambda_{1k_1}$, and $\lambda_{0k_2}\approx
-\lambda_{1k_2}$ (see Sec. IV B). Therefore, the terms
$\lambda_{0k}\lambda_{1k}/\delta_k$ arising from these states have
opposite signs and cancel each other, so that the contribution of
the doublet $\left\{\left|k_1 \right\rangle, \left|k_2
\right\rangle\right\}$ to $\Lambda_2$ is minimal. To estimate the
contributions from higher states with energies $\varepsilon \ge
-0.05$, we should take into account two circumstances. i) The ODT
matrix elements between logical states and excited states with
equal $l$ decrease with the energy growth whereas the detunings
increase with the energy growth. ii) High-lying states are closely
spaced to each other and the distance between two neighboring
states (not necessarily from same doublet) decrease rapidly with
the energy growth. Thus, one may expect that corresponding terms
in $\Lambda_2$ will either cancel out each other, as it was for
degenerate states, or have insignificant effect on the sum
convergence because of their subsequent reduction versus state
energy. We think that both issues are important and their
cooperative effect takes place, establishing our observation on
the oscillation type of $p_{tr}$. Strictly speaking, for long
times $T$,  when $\Omega _R T >
> 1$, this is not true, because an internal structure of the sum
$p_{tr}$, containing the oscillating terms of different but close
frequencies, causes these oscillations to be averaged out and
$p_{tr}$ tends to its average value $\sim\max \left( {p_{tr} }
\right)/2$.

Utilizing similar arguments, we specify an algorithm of choice of
the pulse frequency and amplitude in off-resonant driving regime.
First of all, we determine the doublet whose states
$\left|{\widetilde{k}}_1 \right\rangle $ and
$\left|{\widetilde{k}}_2 \right\rangle $ are characterized by
sufficiently large ODT matrix elements and high tunneling rates
(large value of the energy splitting in comparison with the pulse
energy bandwidth). Then, by adjusting the pulse frequency to the
center of energy gap between doublet states, we obtain for
detunings $\delta _{{\widetilde{k}}_1 }  = - \delta
_{{\widetilde{k}}_2 } $. In this case, the terms ${{\lambda _{0k}
\lambda _{1k} } \mathord{\left/
 {\vphantom {{\lambda _{0k} \lambda _{1k} } {\delta _k }}} \right.
 \kern-\nulldelimiterspace} {\delta _k }}$ enter into $\Lambda_2$ with the same sign
resulting in amplification in the sum. Doing so, we should
remember that the field energy $\varepsilon_{field}$ (i.e., pulse
strength) must be much smaller than the doublet energy splitting.
As it can be seen from Fig. 2 (b), upper bound value for
$\varepsilon_{field}$ is to be set $\sim 0.005$ at the whole
considered range of $R$. Actually, it means that we can reliably
operate with frequencies that fall within the interval containing
the states from HMI subbands with $n=3$ and $n=4$. Moreover, the
frequency choice just outlined cannot guarantee low population of
excited states. A general way to minimize this population implies
the use of moderate pulse strengths at fixed detunings. Note, that
two-level Rabi oscillation regime with $\max(p_{tr})<0.01$ can be
achieved under this frequency choice, since marked points on Fig.
7 lie in the middles of energy gapes separating the doublet states
$\left| 5g\sigma_g \right\rangle$ - $\left| 6h\sigma_u
\right\rangle$, $\left| 6g\sigma_g \right\rangle$ - $\left|
7h\sigma_u \right\rangle$, and $\left| 7i\sigma_g \right\rangle$ -
$\left| 8j\sigma_u \right\rangle$. With that, the pulses have to
be strong enough in order to perform qubit rotations in the times
that are shorter than decoherence time. For the off-resonant
driving scheme, where the decoherence effects due to the finite
population of excited states (e.g., the electron-phonon
relaxation) are reduced considerably, the dephasing of localized
states is expected to be the main source of coherence losses. In
our simulations we require that the operation times ($T_{op}\sim
1/\Omega_R$) do not exceed $10^{-8}$ s. It gives a lower bound
value on the pulse strength to be $\varepsilon_{field}\sim0.001$.
For Si:P$^+_2$ structure parameters, the allowed range
$0.001<\varepsilon_{field}<0.005$ corresponds to the pulse
strengths of hundreds of V/cm that is by two orders smaller than
the field strengths considered in the voltage-based driving
schemes \cite{12} - \cite{16}.

Provided that the pulse parameters are selected correctly, the
probability amplitudes $c_0$ and $c_1$ (and, consequently,
probabilities $p_0$ and $p_1$) evolve smoothly against time thus
allowing one to continuously rotate qubit state vector along fixed
meridian with $\varphi=\pi/2$ on the Bloch sphere, namely, to
create the superposition of logical states of the form $\left|
{\Psi \left( t \right)} \right\rangle  = \sqrt {p_0 \left( t
\right)} \left| 0 \right\rangle  - i\sqrt {p_1 \left( t \right)}
\left| 1 \right\rangle $, where $p_0 \left( t \right)+p_1 \left( t
\right)\approx 1$. Control over relative phase between logical
states, needed for creation of an arbitrary superposition state of
the qubit, requires the logical states to be addressed
independently by two different pulses to maintain
$\Lambda_0\ne\Lambda_1$ in Eq. (20). To achieve this goal, one
should break the central symmetry of HMI making use of external
field, thus the important quantum operations given by Eqs. (34) -
(36) will be realized in this case after application of
bichromatic pulse. Besides, there exists a more familiar way
utilizing the adiabatically varied bias field to produce a shift
between the energies of logical states.

\centerline{\bf D. Phase and population dynamics in nonzero-bias
case}

As it was demonstrated in the works \cite{3}, \cite{16},
\cite{20}, and \cite{30}, an uniform electrostatic field,
polarized along $z$ axis, disturbs spatial symmetry of DD
structure relative to the origin and produces a necessary energy
shifts. The Hamiltonian of HMI, subjected to the action of an
external axial field, reads $H = H_0 - eE_0 z$. We compute the
eigenfunctions $\left\{ {\left| {\tilde \Psi _i } \right\rangle }
\right\}$ of HMI with the field, expanding them over the
eigenfunctions $\left\{ {\left| {\Psi _k } \right\rangle }
\right\}$ without the field: $\left| {\tilde \Psi _i }
\right\rangle = \sum\limits_k {\tilde C_{ki} \left| {\Psi _k }
\right\rangle } $. The eigenenergies and the expansion
coefficients $\tilde C_{ki}$ are calculated directly through the
diagonalization of the Hamiltonian in the basis of 64
$\sigma$-states, found in Sec. IV A. Similarly, the ODT matrix
elements $\left\{ \tilde d_{ij} \right\}$ between the HMI
eigenstates in the non-zero bias case are expressed via the ODT
matrix elements between the HMI eigenstates in the zero bias case,
$\left\{d_{mn} \right\}$, as follows: $\tilde d_{ij} =
\sum\limits_{m,n} {\tilde C_{mi}^* \tilde C_{nj} d_{mn} }$.

The dependencies of eigenenergies on static field energy
$\varepsilon^{st}_{field}$ (defined in complete analogy with
$\varepsilon_{field}$) are given at Fig. 8 for internuclear
distance $R=38$. With respect to the hybridization degree of HMI
states at $\varepsilon^{st}_{field}=0$, the important difference
in energy behavior of states, originated from localized and
delocalized states, is observed at $\varepsilon^{st}_{field}\ne0$.
We see that the energies of states, corresponding to weakly
hybridized states of HMI without bias field (e.g., the states from
subbands with $n=1$ and $n=2$), are approximately linear functions
of $\varepsilon^{st}_{field}$ (due to Stark effect). The
generation of the energy difference $\Delta=\varepsilon_1-
\varepsilon_0=\varepsilon^{st}_{field}R$, where $R=38$, requires
an application of bias field with the strength $E_0 \sim 10$ V/cm
(see Fig. 8 (a)). If $\varepsilon^{st}_{field}>0$, the states,
whose energies bring down (up) with the static field strength, are
localized predominantly on "deep" donor $B$ ("shallow" donor $A$).
From Fig. 8 (b) one can observe a general tendency of the HMI
spectrum to become complicated with the field growth due to both
the full lifting of Coulomb degeneracy (in contrast to the partial
lifting in HMI spectrum without field) and the hybridization of
states pertaining to different subbands (giving rise to
crossing-anticrossing pattern on Fig. 8 (b)).
 Otherwise, the
energies of delocalized states $\left| 5g\sigma \right\rangle$ and
$\left| 6h\sigma \right\rangle$ (thick red curves on Fig. 8(b))
vary with static field energy as $\varepsilon \left[ {5g\left(
{6h} \right)\sigma } \right] \approx \frac{1}{2}\left[
{\varepsilon \left( {5g\sigma _g } \right) + \varepsilon \left(
{6h\sigma _u } \right)} \right] \mp \frac{1}{2}\sqrt {\left(
{\varepsilon _{field}^{st} R} \right)^2 + \Delta _{5g\sigma _g -
6h\sigma _u}^2 } $ , where $\Delta _{5g\sigma _g - 6h\sigma _u}  =
\varepsilon \left( {6h\sigma _u } \right) - \varepsilon \left(
{5g\sigma _g } \right)$ is the tunnel splitting at zero bias
field. It means that the tunneling rate for those states remains
sufficiently high if $\left| {\varepsilon _{field}^{st} } \right|
\le 0.001$. Therefore, we shall continue to operate with those
states exploiting them as the transport channels. The dependencies
of the matrix elements on the field strength for transitions,
connecting logical states and transport states $\left| 5g\sigma
\right\rangle$ and $\left| 6h\sigma\right\rangle$, are given on
Fig. 9. The examination of the values of matrix elements confirms
the fact that the maximum of electron density for the  state
$\left| 5g\sigma \right\rangle$, originated from symmetric state
$\left| 5g\sigma_g \right\rangle$, is displaced onto "deep" donor
$B$, whereas for the state $\left| 6h\sigma \right\rangle$,
originated from asymmetric state $\left| 6h\sigma_u
\right\rangle$, it goes onto "shallow" donor $A$. The deeps on the
curves appear due to the electron density redistribution at
crossing/anticrossing points, where transport states interact with
localized states (see Fig. 8 (b)).

In Ref. \cite {17}, it was demonstrated that elementary
single-qubit operations can be performed in asymmetric DD
structure, driven by two resonant pulses. Here we check this
proposal for zero-detuning case by the simulation of NOT
operation. According to our previous results, the resonant
asymmetric three-level scheme requires that $|\lambda_{0
tr}|=|\lambda_{1 tr}|$, or $|E_0/E_1|=|d_{1 tr}/d_{0 tr}|$. Since
the ODT matrix elements for distinct arms of the excitation scheme
differ from each other, we need to compensate the difference in
coupling coefficients $\lambda_{0 tr}$ and $\lambda_{1 tr}$ by
choosing the pulse strengths so that to fulfill above condition.
Now we integrate Eq. (6) with the right-hand side defined by the
set $\left\{\tilde d_{ij}\right\}$ of the ODT matrix elements and
with the initial conditions reflecting the localization of
electron  at the beginning of pulse action in the ground state
$\left|0 \right\rangle$. As usual, the probabilities $p_0(T)$,
$p_1(T)$, and $p_{tr}(T)$ (Fig. 10) illustrate the qubit state
inversion for the case when both pulses are in the exact resonance
with the state $\left| 5g\sigma \right\rangle$ (the other
parameters are indicated at the plot). We can reveal that the arms
of excitation scheme remain inequivalent, despite of that the
couplings have been balanced. The time dependency of $p_1$ is
smooth, but the plot for $p_0$ demonstrates a fringes superimposed
on ideal "three-level" curve (compare with that pictured on Fig.
5) and originated from non-resonant excitation of "deep" donor
states whose energies lie closely to the $\left| 5g\sigma
\right\rangle$ states. For $\varepsilon^{st}_{field}$=0.0004, the
difference between energy of transport state and those of the
states of donor $B$ is $\sim 0.006$, whereas the energies of
states localized on donor $A$ lie by $\sim 0.02$ higher than the
transport state energy. Add, that the character of dynamical
picture is not changed noticeably, if one uses the state $\left|
6h\sigma \right\rangle$. To suppress these fringes and refine
oscillation picture, one should apply the pulses with lower
strengths. In spite of presence of these perturbations, the
electronic transfer probability is very high and the period of the
oscillations is short enough. We expect this type of evolution to
be conserved for various quantum operations carried out in
asymmetric DD structure under the resonant excitation regime.

Unlike in the case of symmetric HMI, here we were unable to find
pulse parameters for asymmetric off-resonant driving scheme that
would provide a robust and fast implementation of basic quantum
operations. This lack takes place because the pulse with frequency
$\omega_0$, being tuned slightly below the state $\left| 5g\sigma
\right\rangle$ for the transition $\left|
0\right\rangle\leftrightarrow\left| 5g\sigma \right\rangle$,
addresses unwanted nearly-resonant transitions connecting the
state $\left| 1\right\rangle$ with the states localized on
"shallow" donor. Thus, the selectivity requirement is violated.
The attempts to achieve the reliable two-level oscillation picture
have forced us both to reduce the pulse strengths and to vary the
detuning in such a way that both pulses would drive only the
transitions prescribed by ideal excitation scheme. The goal is
likely to be attained by this method, but the characteristic times
will be inappropriately long in comparison with decoherence time.
In particular, the asymmetric off-resonant scheme, described in
Sec. III, will function correctly if $\Delta$ is much smaller than
the energy spacings between the states lying around the energy
$\varepsilon_0+\omega_0$. At same time, the coupling coefficients
must be much smaller than $\Delta$, in order to address separately
each arm of driving scheme by its own pulse. From the Fig. 8 (b)
we obtain the estimations $\varepsilon^{st}_{field}<10^{-4}$ and
$\Delta<0.004$ that, in its turn, implies $|\lambda_{0(1)k}|$ to
be less than $10^{-3}$. It tells us that under these conditions
the off-resonant Rabi frequency $\Omega_R$ is of order of $10^{8}$
s$^{-1}$ (or smaller) that is out of interest of this paper.

Actually, it means that we can only deal with symmetric DD
structure where the possibility of realization of the off-resonant
qubit-state $\sigma_X$ rotation has been already demonstrated.
However, we know this type of quantum evolution is not sufficient
for implementation of arbitrary quantum operation, since the phase
control is also required. This difficulty may be overcome with the
help of direct introduction of the phase difference between
logical states by the voltage pulse that produces necessary energy
shift. This technique is widely used in quantum dot structures to
align the energy levels of different dots that is needed for
experimental investigations of their transport and spectral
properties. Here we consider a simplified model of voltage pulse
action on the HMI replacing non-isotropic electric field,
generated by the gate $V_a$ (Fig. 1 (a)), by uniform axial
electric field, as it is demonstrated at Fig. 1 (b). The time
dependency of voltage pulse is approximated by the step function.
In this case, elementary calculations predict the qubit phase
oscillations with the frequency $\Omega_{phase}=\Delta/\hbar$. The
general expression for qubit-state evolution operator in the
two-level approximation, accounting for different voltage pulse
shapes, can be found in Ref. \cite{30}. We illustrate this type of
quantum dynamics at Fig. 11, where two-level Rabi oscillations
between the states $\left| + \right\rangle _H  = {{\left( {\left|
0 \right\rangle  + \left| 1 \right\rangle } \right)}
\mathord{\left/
 {\vphantom {{\left( {\left| 0 \right\rangle  + \left| 1 \right\rangle } \right)} {\sqrt 2 }}} \right.
 \kern-\nulldelimiterspace} {\sqrt 2 }}$ and
$\left|  -  \right\rangle _H  = {{\left( {\left| 0 \right\rangle
- \left| 1 \right\rangle } \right)} \mathord{\left/
 {\vphantom {{\left( {\left| 0 \right\rangle  - \left| 1 \right\rangle } \right)} {\sqrt 2 }}} \right.
 \kern-\nulldelimiterspace} {\sqrt 2 }}$ obtained from logical states by the Hadamard rotation,
are presented. Relative to computational basis, these oscillations
indicate on the phase dynamics generated by the operator
$\sigma_Z$. For example, the application of voltage pulse during
one-half oscillation period amounts to the phase shift by $\pi$
between the states $\left|0 \right\rangle$ and $\left|1
\right\rangle$ and takes about several picoseconds. With that, the
population leakage from logical subspace remains negligible
($p_{tr}<10^{-6}$). At the intersection points, the Hadamard
superpositions $\left|  \pm  \right\rangle _H$ collapse into one
of logical states.

Summarizing the results obtained in this and previous subsections,
we conclude the numerical study of electron dynamics by
observation that the strategy, utilizing sequential implementation
of $\sigma_X$ and $\sigma_Z$ rotations, appears to be more
reliable than the strategy based on simultaneous action of voltage
and optical pulses. With that, one may regard the formulas deduced
in Sec. III for asymmetric off-resonant scheme as describing
non-ideal nearly-symmetric scheme driven by single pulse, where
other external fields (e.g., acting on nearest qubits) slightly
disturb the spatial symmetry of structure in uncontrollable way.
In this case, small differences in frequencies of the arms
quantified by the asymmetry parameter $\Delta$, will amount to
quantum error.

Of course, we should take into account that the driving scheme of
Eqs. (34) - (36), whose efficiency has not been confirmed above
for effective HMI, can describe the qubit dynamics for other
models of DD structure (for example, that of heteropolar molecular
ions \cite{16}). The main issue, that makes the verification of
proposed algorithm of quantum control difficult, is concerned with
necessity to know in details the energy spectrum and the ODT
matrix elements for each structure under consideration.

\centerline{\bf V. DISCUSSION }

To provide more clarity in the understanding of the advantages of
the charge qubit-state engineering presented above, let us compare
the resonant and off-resonant excitation schemes. The resonant
optical driving of the DD structure modelled by single three-level
scheme has been studied in Refs. \cite{17,18}. In the case
considered here a more complex dynamics can take place involving
more than one three-level scheme. For example, if we tune the
lasers on resonance with the transition between the logical
subspace $\left\{ {\left| 0 \right\rangle ,\left| 1 \right\rangle
} \right\}$ and a state $\left| {r} \right\rangle ,\,\,r \in
\left\{ {k } \right\}_{exc}$ located near the top of the barrier,
a number of states with the energies close to $\varepsilon _{r} $
will be excited as well. This picture is quite expected in the
hydrogen-like molecular ions for the high-lying states which
energies are within the interval $\Delta \varepsilon _{r} \le
\left| {\lambda _{0(1)r} } \right|$. However, for the symmetric
structure it means that there will be no transitions between the
qubit states. It is because the exited states belonging to the
same doublet are presented by the symmetric and antisymmetric
superpositions of the excited states of isolated donors which,
 being excited simultaneously, interfere constructively on one donor
  and destructively on another one. As a consequence, the donors
are excited independently. This effect becomes more significant as
the interdonor distance $R$ increases and the tunnel coupling
between the donors decreases. We have observed such type of
electronic dynamics when pulse was tuned into resonance with
low-lying excited states with $l=s,p,d,f$. When the energy
splitting of the maximally-resolved doublet becomes comparable
with the coupling coefficients of the optical dipole transitions,
the process of the electron transfer between the donors is
terminated. Note that the optically driven DD structure will
demonstrate the similar behavior if one of the pulses is short
enough so that its duration is $T \le {1 \mathord{\left/
 {\vphantom {1 {\Delta \varepsilon _{r} }}} \right.
 \kern-\nulldelimiterspace} {\Delta \varepsilon _{r} }}$ and thus
 it contains harmonics in the frequency range $\delta \omega  \sim {1 \mathord{\left/
 {\vphantom {1 T}} \right.
 \kern-\nulldelimiterspace} T} \ge \Delta \varepsilon _{r} $.
 Again, the states with the energies belonging to the interval $\Delta \varepsilon _{r}$ will be
 excited simultaneously giving rise to the electron transfer
 blockage just outlined. In our simulations, we have arrived at
 this regime via application to HMI a short and intense
 ($\varepsilon_{field}\sim0.1$) pulse.

The reliable resonant scheme thus deals with single transport
state (for HMI, the states $\left|5g\sigma_{(g)} \right\rangle$
and $\left|6h\sigma_{(u)} \right\rangle$) and is very sensitive to
the pulse detuning from the resonance with that state. For
example, the non-zero detuning always produces an amplitude error
in NOT gate because of incomplete depopulation of the initial
state when the pulse is off \cite{21}. On the contrary, the use of
the off-resonant pulses enables one to exploit the whole number of
excited states (from which only several ones participate
substantially in electronic dynamics). Moreover, we don't need to
control the pulse frequencies  with high accuracy since a small
variation in the pulse detunings brings about an insignificant
change in the Rabi frequency (see Fig. 7). The computational
errors originated from the frequency renormalization can be
corrected by the corresponding change in the pulse duration due to
the smooth time dependencies of the probability amplitudes, as it
was shown at Fig. 6. The only requirement that must be followed
closely for successful electron state manipulations is the Raman
two-photon resonant condition (9).

The selectivity of the electron resonant transfer requires also a
strict control over the pulse polarizations. The transport states
in the molecular ion are formed through the hybridization of those
individual donor states whose wave functions are extended along
the axis $z$ that coincides with the interdonor direction. Other
states (e.g., $\pi$-states) are hybridized weakly and cannot
assist efficiently in the electron dynamics. Their excitations are
due to the pulse components polarized along the axes $x$ and $y$.
It amounts to the population leakage into the non-hybridized
single-donor states with the energies lying in the close proximity
to the energy of the transport state. Let us define the small
angles $\gamma_{n\,x}$ and $\gamma_{n\,y}$ that characterize the
deviations of the $n$-th pulse polarization from the axis $z$:
\begin{equation}
\begin{array}{l}
 {\bf{E}}_n  = {\bf{E}}_{n\,z} + {\bf{E}}_{n\,x} \cos \left( {{\pi  \mathord{\left/
 {\vphantom {\pi  2}} \right.
 \kern-\nulldelimiterspace} 2} + \gamma _{n\,x} } \right) + {\bf{E}}_{n\,y} \cos \left( {{\pi  \mathord{\left/
 {\vphantom {\pi  2}} \right.
 \kern-\nulldelimiterspace} 2} + \gamma _{n\,y} } \right), \\
 \,\,\,\,\,\,\,\,\,\,\,\,\,\,\,\,\,\,\,\,\,\,\,\,\,\,\,\,\,\,\,\,\,\,\,\,\,\,\,\,\,\,\,\,\left| {\gamma _{n\,x} } \right|,\,\,\left| {\gamma _{n\,y} } \right| \ll 1,\,\,\,\,\,n = 0,1 ,\\
 \end{array}
\end{equation}
 then the probability of successful implementation of the quantum
 operations is reduced by a factor of $w \sim 1 - \max({\gamma _ {nx}^2,\gamma _ {ny}^2})$. In the off-resonant case, the
populations of those states remain negligibly small ($ \sim
{{\left| {\lambda _{n k} } \right|^2 } \mathord{\left/
 {\vphantom {{\left| {\lambda _{0,1\,k} } \right|^2 } {\delta _k^2 }}} \right.
 \kern-\nulldelimiterspace} {\delta _k^2 }}$) and the
corresponding channel of population leakage is blocked.

The important difference between the resonant and off-resonant
schemes lies in the treatment of the decoherence problem. We know
the relaxation rates from the transport state caused by the
spontaneous photon/phonon emission during the resonant excitation
\cite{18} may be high enough to corrupt the qubit state. In the
off-resonant scheme the population of the intermediate state(s) is
negligible and the probability of relaxation is drastically
reduced. The influence of the residual population of the
intermediate state on the adiabatic electron transfer in the
three-level scheme was examined in Ref. \cite{31} for the gaussian
pulses. It was shown that the error introduced by the spontaneous
emission together with the error due to the non-adiabaticity are
inversely proportional to the pulse detuning and can be made small
enough to allow the fault tolerant quantum computation.

Note that the complete population transfer between the qubit
states, or NOT
 operation, requires that $\Lambda _{0}  = \Lambda _{1}$. This
is naturally met for nearly symmetric DD structures where $\Delta
\approx 0$ and $\left| {{\bf{d}}_{0k} } \right| \approx \,\left|
{{\bf{d}}_{1k} } \right|$. In general, however, one should keep in
mind that $\left| {{\bf{d}}_{0k} } \right| \ne \,\left|
{{\bf{d}}_{1k} } \right|$ that makes the performing of the
condition $\Lambda _{0}  = \Lambda _{1}$ very problematic. It
seems then reasonable to point the other way for the population
transfer based upon the pulse-shaped techniques. Such methods,
e.g., the stimulated Raman adiabatic passage (STIRAP) \cite{26},
are very robust against the pulse/structure imperfections and
would allow one to handle the quantum information carefully. The
theory of the adiabatic population transfer via multiple
intermediate states, including the off-resonant case, was
presented in Ref. \cite{27}. Note that for the pulses strongly
detuned from the resonance, the time ordering is no more important
since successful population transfer may be attained for both
intuitive and counterintuitive pulse sequences. If initially $c_0
\left( t_0 \right) = 1,\,c_1 \left( t_0 \right) = 0$, the
intuitive (counterintuitive) pulse ordering means that $\mathop
{\lim }\limits_{t \to  t_0 } \left[ {{{f_0 \left( t \right)}
\mathord{\left/
 {\vphantom {{f_0 \left( t \right)} {f_1 \left( t \right)}}} \right.
 \kern-\nulldelimiterspace} {f_1 \left( t \right)}}} \right] = \infty \left( 0
 \right)$ and $\mathop {\lim }\limits_{t \to T } \left[ {{{f_0 \left( t
\right)} \mathord{\left/
 {\vphantom {{f_0 \left( t \right)} {f_1 \left( t \right)}}} \right.
 \kern-\nulldelimiterspace} {f_1 \left( t \right)}}} \right] = 0\left( \infty
 \right)$ and, as it follows from Eq. (24), $\Theta \left( t_0 \right) = 0\left( \pi
 \right)$, $\Theta \left( T  \right) = \pi \left( 0 \right)$.
 The population transfer may be understood as the adiabatic temporal development of the
eigenstate $\left| +  \right\rangle$ ($\left| -  \right\rangle$)
for the intuitive (counterintuitive) pulse ordering. As it is seen
from Eq. (32), the qubit state inversion is realized in the
asymmetric DD structures if the conditions $\arg \left[ {\Lambda
_2 \left( T \right)} \right] \pm \tilde \Omega \left( T \right) =
\pi n$ and $T\Delta = \pi \left( {2m + 1} \right)$ are fulfilled.
The detailed analysis concerning the arrangement of the pulse
shapes in STIRAP can be found elsewhere \cite{26}.

The effect of the detuning $\delta_{two-ph}$ from two-photon
resonance (the spacing between dashed horizontal lines on Fig. 1
(b)) should be also taken into account at more profound level of
investigations. In nearly-symmetric off-resonant case
($\left|\Lambda_0-\Lambda_1\right|\ll\left|\Lambda_2\right|$) the
problem enables analytical solution. If electron was initially
localized into the state $\left|0\right\rangle$, the total
excitation probability (including transfer probability into state
$\left|1\right\rangle$) at large detunings ($\left| {\Lambda _2 }
\right| \ll \left| {\delta _{two - ph} } \right|$) is of order of
$\left( {{{\Lambda _2 } \mathord{\left/
 {\vphantom {{\Lambda _2 } {\delta _{two - ph} }}} \right.
 \kern-\nulldelimiterspace} {\delta _{two - ph} }}} \right)^2  \ll 1$. Therefore,
if the detunings from two-photon resonance are significant,
electronic excitations from localized state $\left|0\right\rangle$
do not occur. In the opposite case of small detunings, i.e., when
$\left| {\delta _{two - ph} } \right| \ll \left| {\Lambda _2 }
\right|$, the probability of successful electronic transfer is
given by formula $p_1^{} \approx 1 - \left( {{{\delta _{two - ph}
} \mathord{\left/
 {\vphantom {{\delta _{two - ph} } {2\Lambda _2 }}} \right.
 \kern-\nulldelimiterspace} {2\Lambda _2 }}} \right)^2 $ (provided that other parameters
are chosen in optimal way).

Finally, in our analytical treatment we consider continuum states
only as virtually excited transport channels, neglecting the
possibility of two-photon resonant electron transitions from
logical subspace into the continuum states, that reside within
interval around the energy $\varepsilon _{2\omega_0}  =
\varepsilon _0 + 2\omega_0 $, via intermediate low-lying excited
states (not necessarily highly populated) with energies $\sim
\varepsilon _0 + \omega_0 $. It is known that in isolated hydrogen
atom the matrix elements for the transition $\left| {1s}
\right\rangle \leftrightarrow \left| {2p\sigma } \right\rangle $
and the matrix elements for transitions connecting the state
$\left| {2p\sigma } \right\rangle $ with $s$ (or $d$) continuum
waves with energies pertaining to the interval pointed above, have
the same order of magnitude. Hence, one could expect the electron
dynamics in DD structure induced by optical driving pulses to be
much more complex. However, in experimental investigations of
low-frequency excitations in atomic systems these processes do not
manifest themselves at the level that would establish the
necessity to include the continuum into theoretical model.
Perhaps, it could be explained by the arguments we have used in
Sec. IV C to account for our numerical results reflecting very
small influence of high-lying states of discrete part of HMI
spectrum on electron dynamics. In principle, the continuum states
being addressed directly can also be used as transport channels,
but, at the same time, can bring about additional decoherence
(see, e.g., \cite{32}).

\centerline{\bf VI. CONCLUSIONS }

In this paper we have considered the one-electron double-donor
structure subjected to the action of optical and electrostatic
pulses. Unlike the other systems proposed to serve as the
potential candidates for the solid state optically-controlled
qubits (double quantum dots, rf-SQUIDs), the double-donor
structure is characterized by sufficiently high density of the
bound states at the edge of the barrier that separates the donors.
It means that the three-level resonant scheme proposed earlier to
implement the desired qubit-state evolution may be unsuitable to
maintain the appropriate selectivity of the optical excitations.
On the other hand, the off-resonant scheme looks as more efficient
for the qubit manipulations and robust in comparison with the
resonant scheme. Though the Raman evolution of the qubit is slower
than that in the case of the resonant driving, it seems to be more
reliable for the implementation of quantum operations. We have
shown that the basic single-qubit operations may be performed on
the DD structure for several pulse and structure parameter
choices. Numerical simulations, carried out on structure modelled
by effective hydrogen molecular ion, have confirmed the validity
of our analytical framework where three important simplifications,
concerning the treatment of dynamical problem, have been made.
There are: i) rotating-wave approximation, ii) adiabatic
elimination, and iii) neglecting the transitions between
high-lying states. Although all of them are widely used, it is, to
our knowledge, for the first time when these approximations are
verified in rather complicated dynamical study of multilevel
system.

The information about the structure and pulse parameters is
contained in the Rabi frequency of the two-level oscillations.
This frequency can be defined experimentally for each set of the
detunings, the strengths, and the durations of the pulses. The
results of those measurements could be used to reconstruct the
features of the spectrum of the DD structure.

Note that the method of the electron-state manipulations by
optical means can be applied also to the spin-based encoding
schemes like that of Ref. \cite{2}. The implementation of
optically controlled effective electron spin exchange described in
Ref. \cite{33} for the two-electron double-dot structure, can be
generalized on the two-electron DD structure. The use of the
adiabatic schemes is of the particular interest.

\centerline{\bf ACKNOWLEDGMENTS }

Discussions with L. A. Openov are gratefully acknowledged.

\newpage

\vskip 6mm

\centerline{\bf FIGURE CAPTIONS }

Fig. 1 (color online). a) Schematics of the quantum state
manipulation in the one-electron double-donor structure. A pair of
(phosphorous) donors $A$ and $B$ (one of them being
singly-ionized) implanted into semiconductor (Si) matrix is
addressed by optical pulse(s). Additionally, electrostatic pulse
generated by voltage gate $V_a$ varies the structural potential.
The desired final orbital state of electron is attained due to
cooperative effect of both pulses on the structure. b) Potential
profile of DD structure, modelled by effective hydrogen molecular
ion (see Sec. IV), along structural axis $z$. The qubit states
$\left| 0 \right\rangle $ and $\left| 1 \right\rangle $ are
defined by the localized orbital states of the donors $A$ and $B$
with the energies $\varepsilon _ 0$ and $\varepsilon _ 1$,
respectively. They are coupled to a collection of excited states
by two optical pulses with the frequencies $\omega_0$ and
$\omega_1$. Uniform axial electrostatic field is applied across
the structure in order to break the symmetry and to introduce the
energy difference $\Delta=\varepsilon _ 1 - \varepsilon _ 0$.
Here, the distance $R$ between donor centers is 38 a.u. and the
electrostatic field energy
$\varepsilon^{st}_{field}=4\times10^{-4}$ a.u. so that
$\Delta=R\times\varepsilon^{st}_{field}=0.0152$ a.u. The energies
of the whole molecular ion are obtained from electronic
eigenenergies + energy of internuclear repulsion $1/R$.

Fig. 2 (color online). The dependencies of electronic
eigenenergies of 20 low-lying eigenstates from $\sigma$-subspace
($m=0$) of hydrogen molecular ion on the internuclear distance
$R$. a) Ground-state energies $\varepsilon(1s\sigma_g)$ and
$\varepsilon(2p\sigma_u)$. b) Excited-state energies. Here, the
eigenenergies of transport states $\left| 5g\sigma_g\right\rangle$
and $\left| 6h\sigma_u \right\rangle$ are presented by thick red
curves whereas the eigenenergies of the states which may also be
used as transport ones are drawn by thick blue lines. The vertical
dotted lines correspond to the internuclear separations $R=22$
a.u., $R=30$ a.u., and $R=38$ a.u., for which dynamical
simulations are performed.

Fig. 3 (color online). Characteristic electronic tunneling time
$\tau _{1s\sigma _g  - 2p\sigma _u }$ between the localized states
$\left| 1s\sigma_g\right\rangle$ and $\left| 2p\sigma_u
\right\rangle$  of hydrogen molecular ion, calculated for
parameters that correspond to single-valley approximation for
Si:P$^+_2$ double-donor structure, as function of the internuclear
distance $R$.

Fig. 4 (color online). The values of matrix elements of optical
dipole transitions, connecting the logical state $\left|
0\right\rangle$ and the low-lying excited states pertaining to the
hydrogen molecular ion subbands with a) $n=2$, b) $n=3$, and c)
$n=4$, as functions of the internuclear distance $R$.

Fig. 5 (color online). An example illustrating the resonant
qubit-state manipulation. The time dependencies of probabilities
$p_0 (T)$ and $p_1 (T)$ to find electron into logical states
$\left| 0\right\rangle$ and $\left| 1 \right\rangle$ and that of
total probability $p_{tr} (T)$  to find electron into excited
states are plotted for initial condition $c_0 (0)=1$, $c_{k\ne0}
(0)=0$. The numerical solutions are given by solid curves, and the
analytical solutions, presented by equation (40) and supplied with
primes at the insets, are visualized by dashed curves. The pulse
is in exact resonance with transport state $\left|
6h\sigma_u\right\rangle$. Other parameters are given at the plot.

Fig. 6 (color online). An example illustrating the off-resonant
qubit-state manipulation. The probabilities $p_0 (T)$, $p_1(T)$,
and $p_{tr}(T)$  demonstrate almost ideal two-level oscillation
picture. The maximum of total probability of population leakage
from computational subspace is 0.01. The pulse frequency is tuned
into the middle of HMI subband with $n=4$ (red-squared point on
Fig. 7 (b)). Other parameters are given at the plot.

Fig. 7 (color online). The Rabi frequency $|\Lambda_2|$ for the
off-resonant symmetric driving scheme is plotted vs the pulse
frequency $\omega\equiv\omega_0$ at two values of pulse strength
$\varepsilon_{field}=0.003$ a.u (blue). and
$\varepsilon_{field}=0.005$ a.u. (green) for internuclear
distances a) $R=30$ a.u. and b) $R=38$ a.u. The curves correspond
to analytical results of Eq. (21) whereas full circles mark the
values of $|\Lambda_2|$ extracted from numerical data. Rabi
frequencies, at which the probability $p_{tr}$ of leakage from
logical subspace is lower than 0.01, are enclosed into red open
squares. Vertical dotted lines designate the energy levels near
which off-resonant approximation becomes inapplicable.

Fig. 8 (color online). The electronic eigenenergies of HMI plotted
in the dependence on static field energy
$\varepsilon^{st}_{field}$ for $R=38$ a.u. Dotted vertical line
designates the value $\varepsilon^{st}_{field}=4\times10^{-4}$
a.u. for which the dynamical simulation is carried out. a) The
energies $\varepsilon_0$ and $\varepsilon_1$ of logical states. b)
The energies of excited states. Thick red curves denote the
eigenenergies of doublet states $\left| 5g\sigma\right\rangle$ and
$\left| 6h\sigma\right\rangle$ that remain delocalized in the
presence of electrostatic field, and can be exploited as transport
channels for resonant manipulations on electronic wave function.

Fig. 9 (color online). The matrix elements of optical dipole
transitions, connecting the logical states $\left| 0\right\rangle$
(dashed curves) and $\left| 1 \right\rangle$ (solid curves) with
the states $\left| 5g\sigma\right\rangle$ and $\left|
6h\sigma\right\rangle$, in the dependence on static field energy
$\varepsilon^{st}_{field}>0$ for $R=38$ a.u.

Fig. 10 (color online). Resonant manipulation of electron orbital
state in asymmetric HMI at the electrostatic field energy
$\varepsilon^{st}_{field}=4\times10^{-4}$ a.u. Both pulses are in
exact two-photon resonance with state $\left|
5g\sigma\right\rangle$ and their amplitudes (field energies
$\varepsilon_{field \,\, 0}$ and $\varepsilon_{field \,\, 1}$) are
chosen so as to equalize the coupling coefficients of effective
three-level excitation scheme (see text for details).

Fig. 11 (color online). The populations $p_+(T)$ and $p_-(T)$ of
Hadamard-rotated states $\left| +\right\rangle_H$ and $\left| -
\right\rangle_H$ vs electrostatic pulse duration $T$. Two-state
Rabi oscillations in the Hadamard frame illustrate dynamics of
relative phase between logical states $\left| 0\right\rangle$ and
$\left| 1 \right\rangle$. The population leakage into excited
states is negligible.

\newpage
\centerline{\includegraphics[width=10cm]{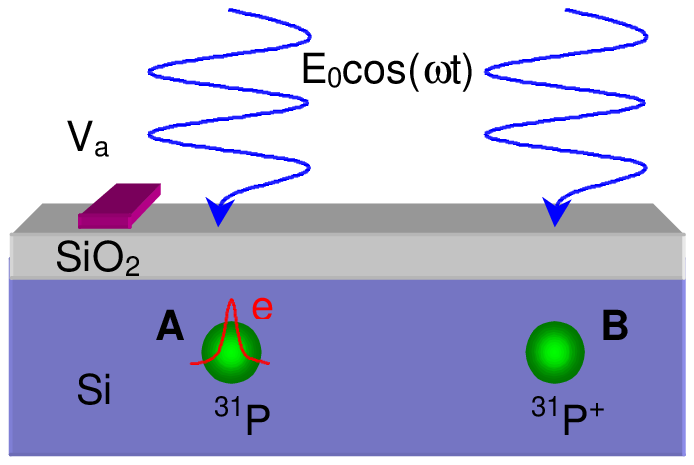}}
 \centerline{Fig. 1 (a)}

\centerline{\includegraphics[width=12cm]{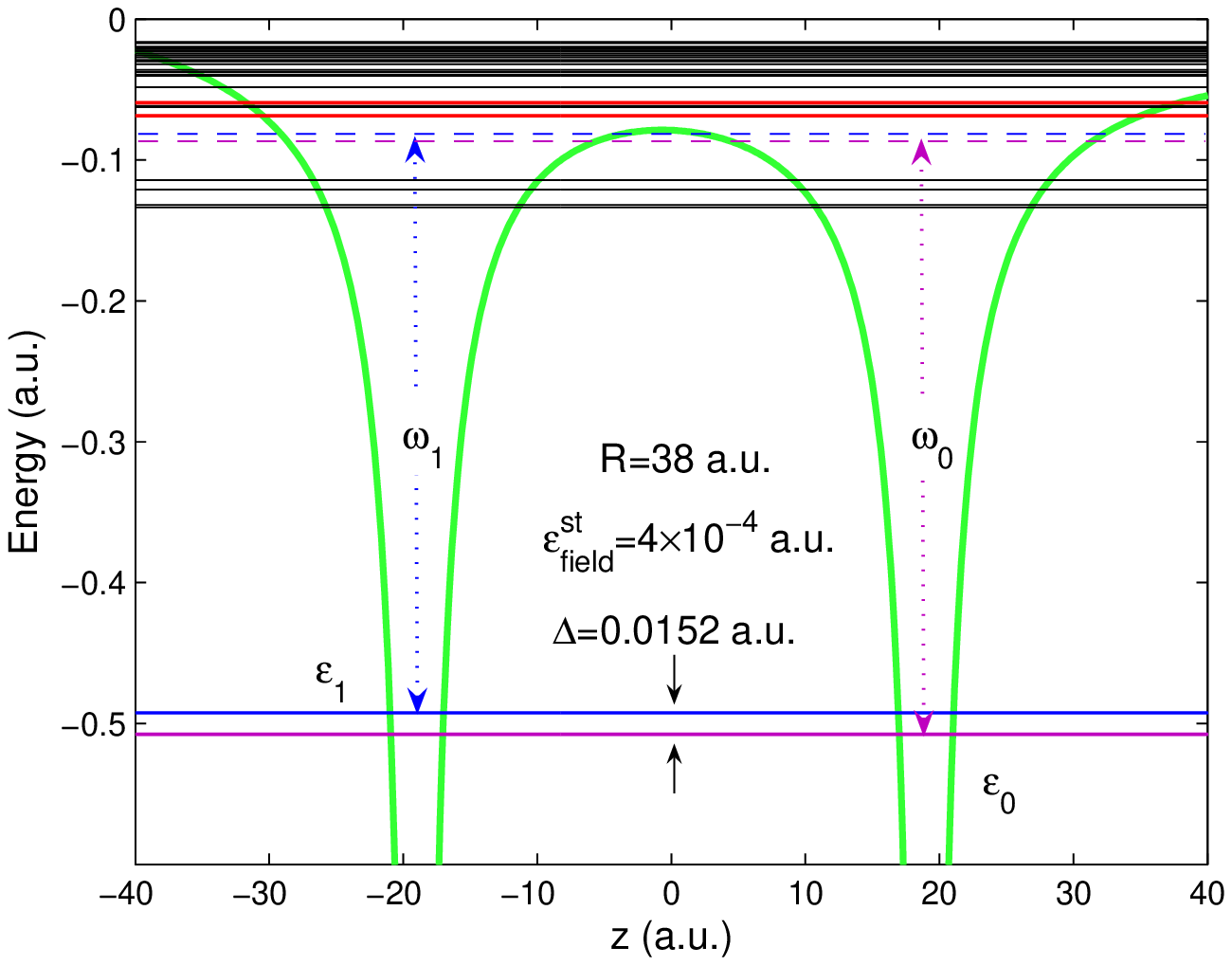}}
 \centerline{Fig. 1 (b)}

\newpage
\centerline{\includegraphics[width=12cm]{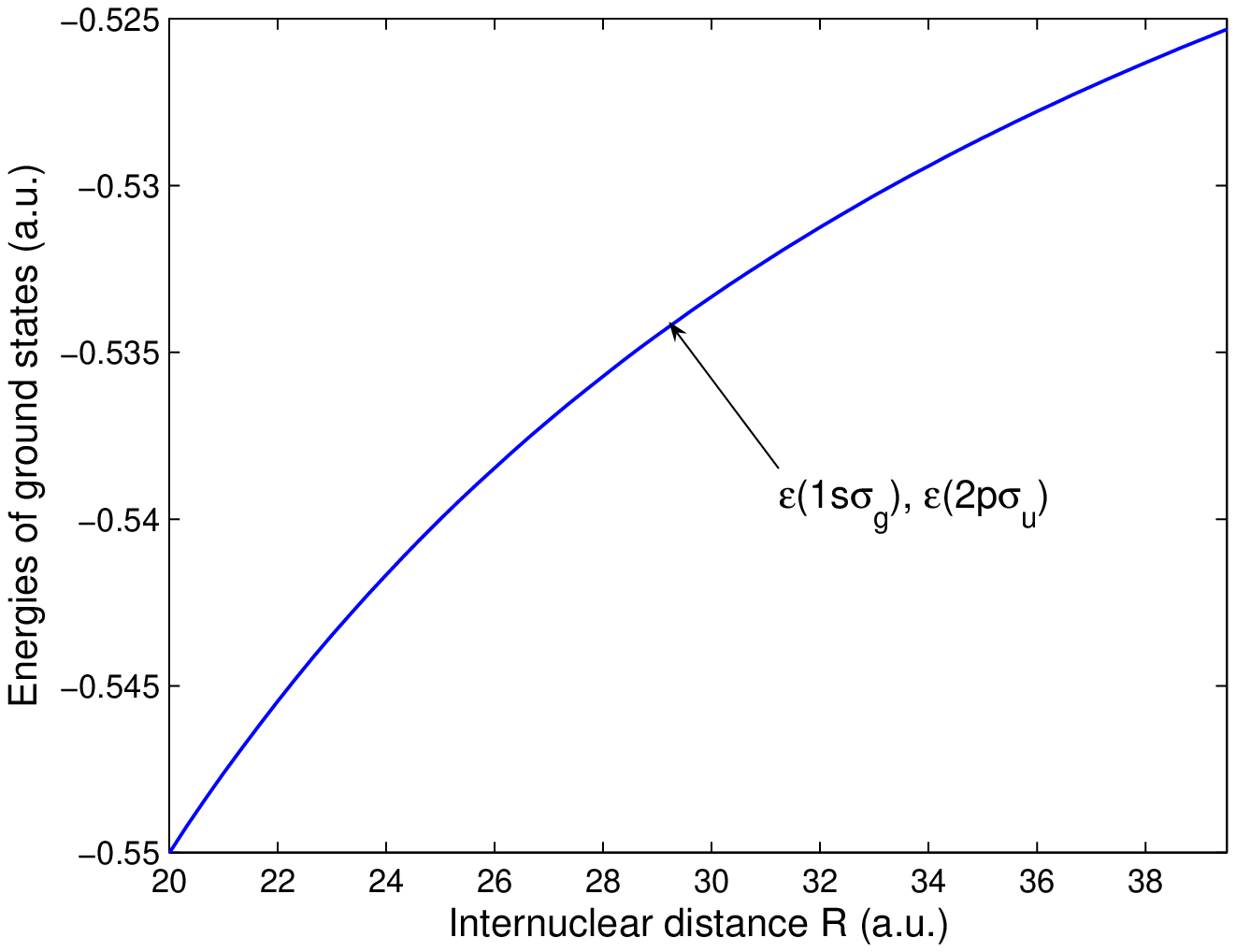}}
 \centerline{Fig. 2 (a)}

\centerline{\includegraphics[width=12cm]{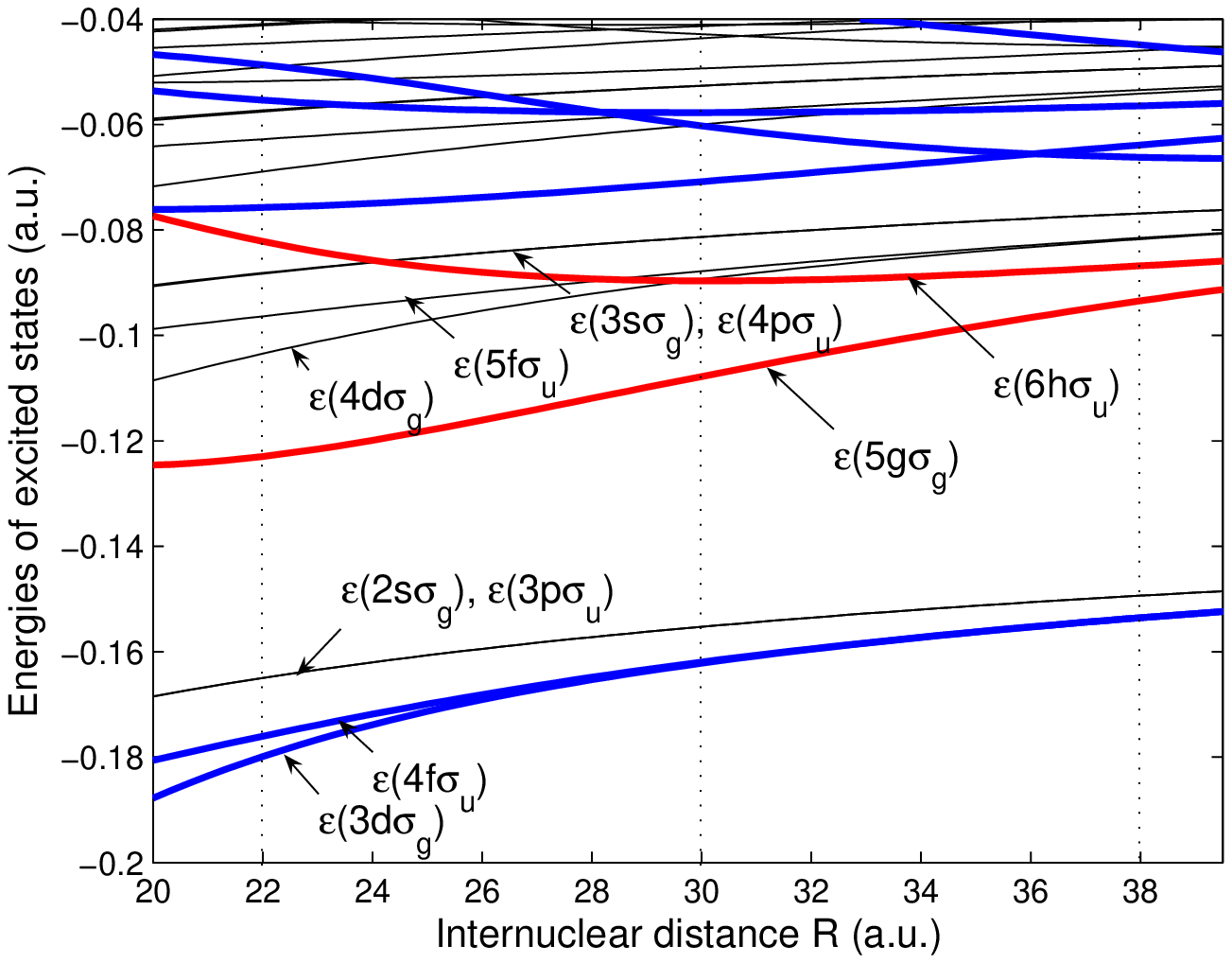}}
 \centerline{Fig. 2 (b)}

\newpage
\centerline{\includegraphics[width=12cm]{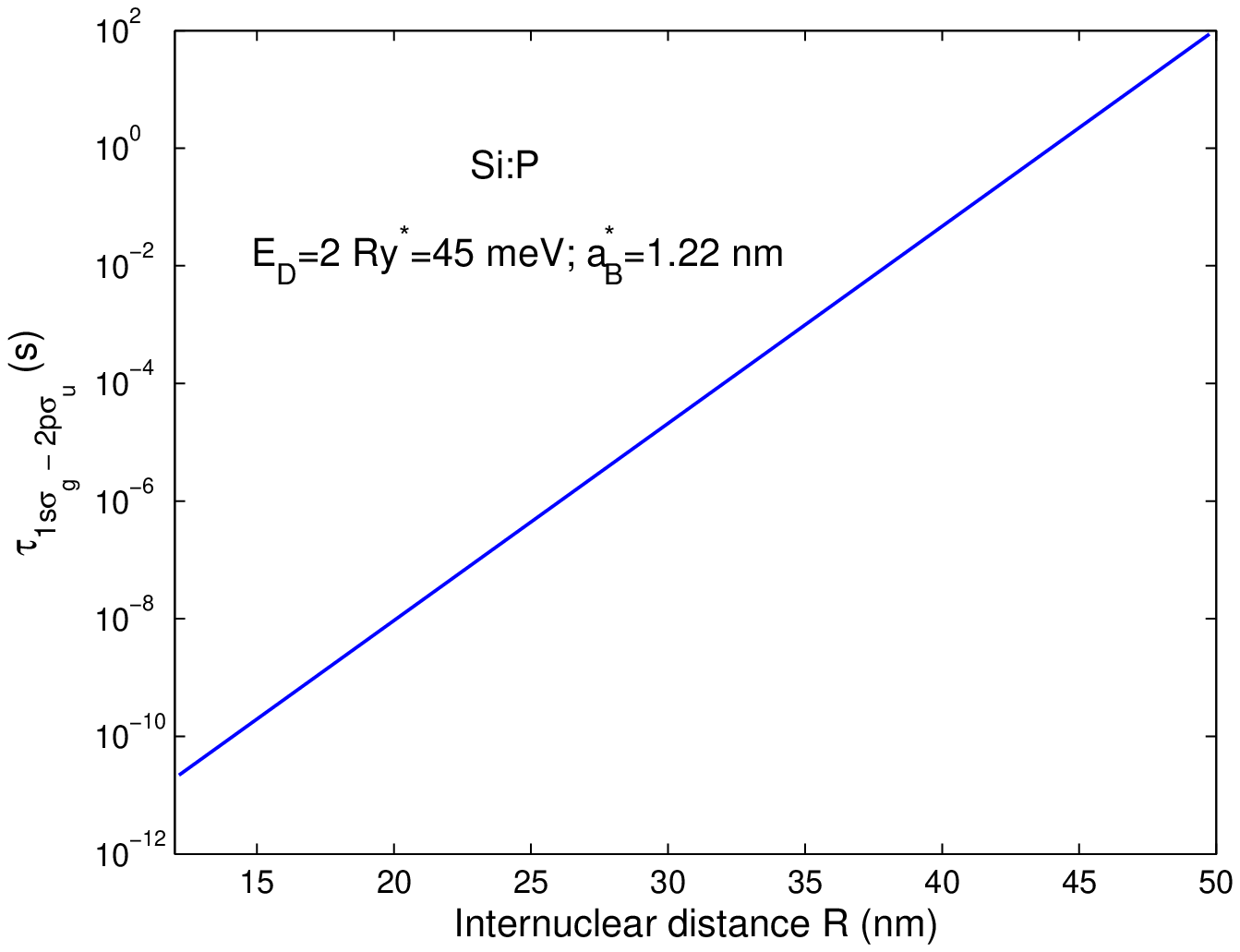}}
 \centerline{Fig. 3}

\centerline{\includegraphics[width=12cm]{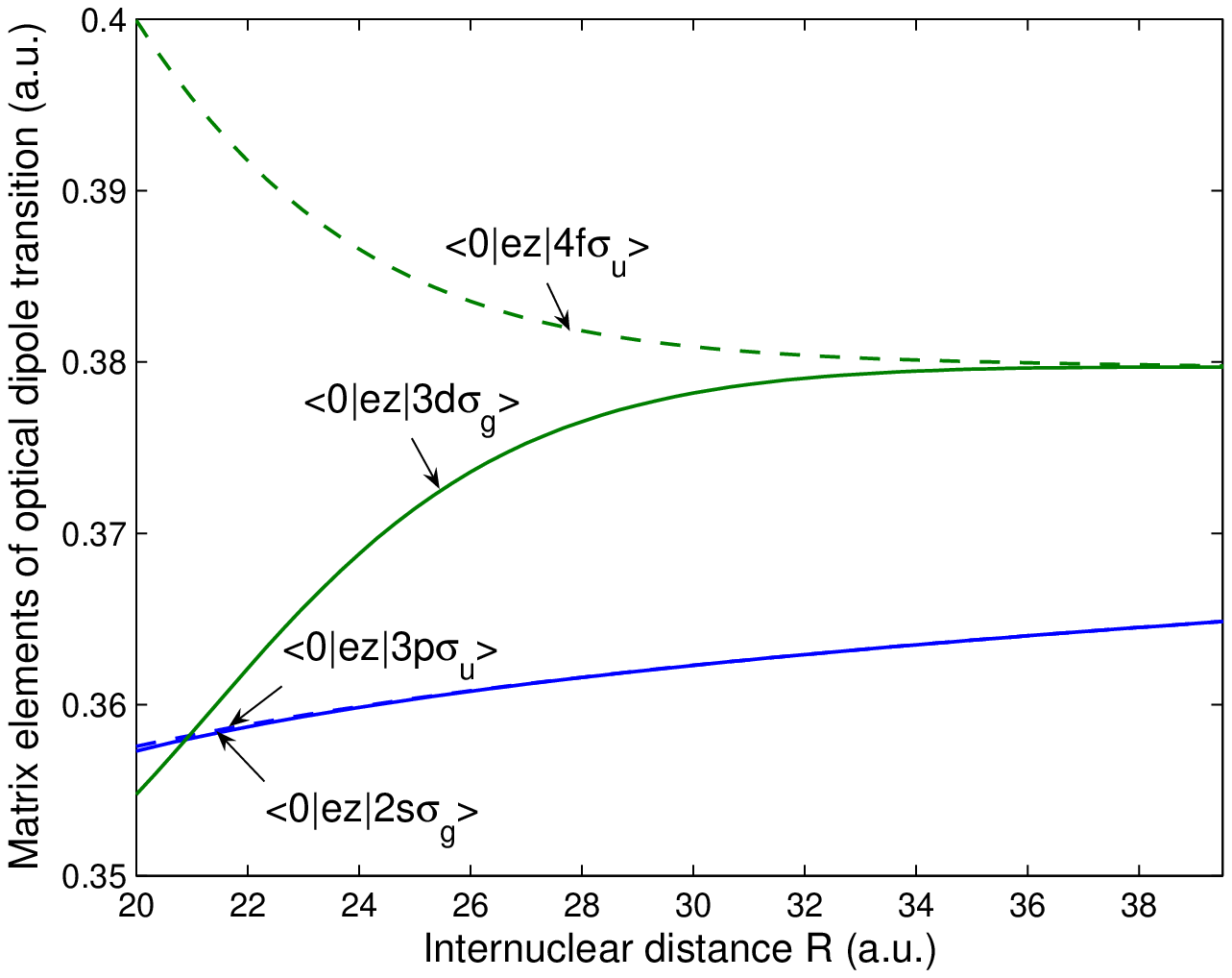}}
 \centerline{Fig. 4 (a)}

\newpage
\centerline{\includegraphics[width=12cm]{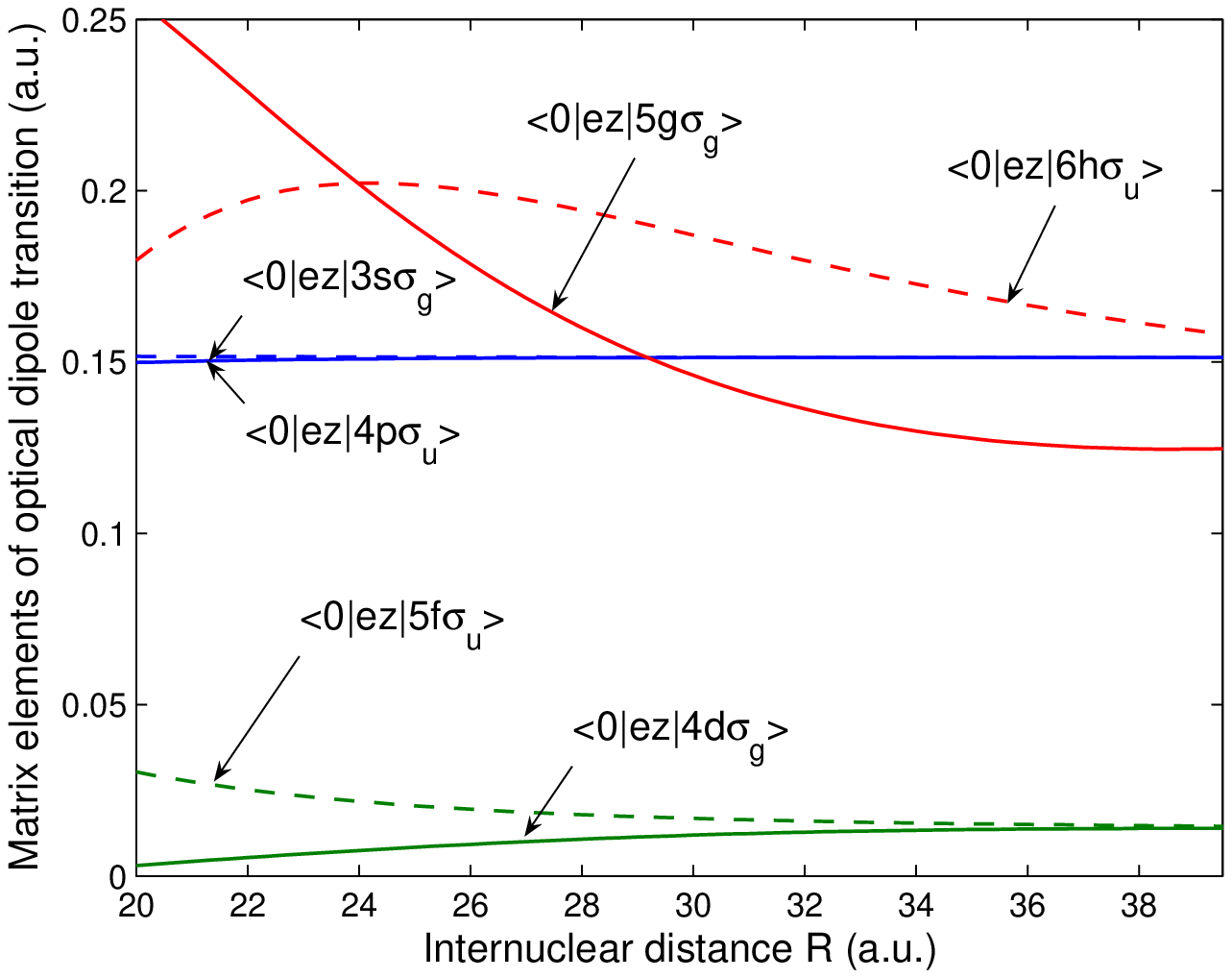}}
 \centerline{Fig. 4 (b)}

\centerline{\includegraphics[width=12cm]{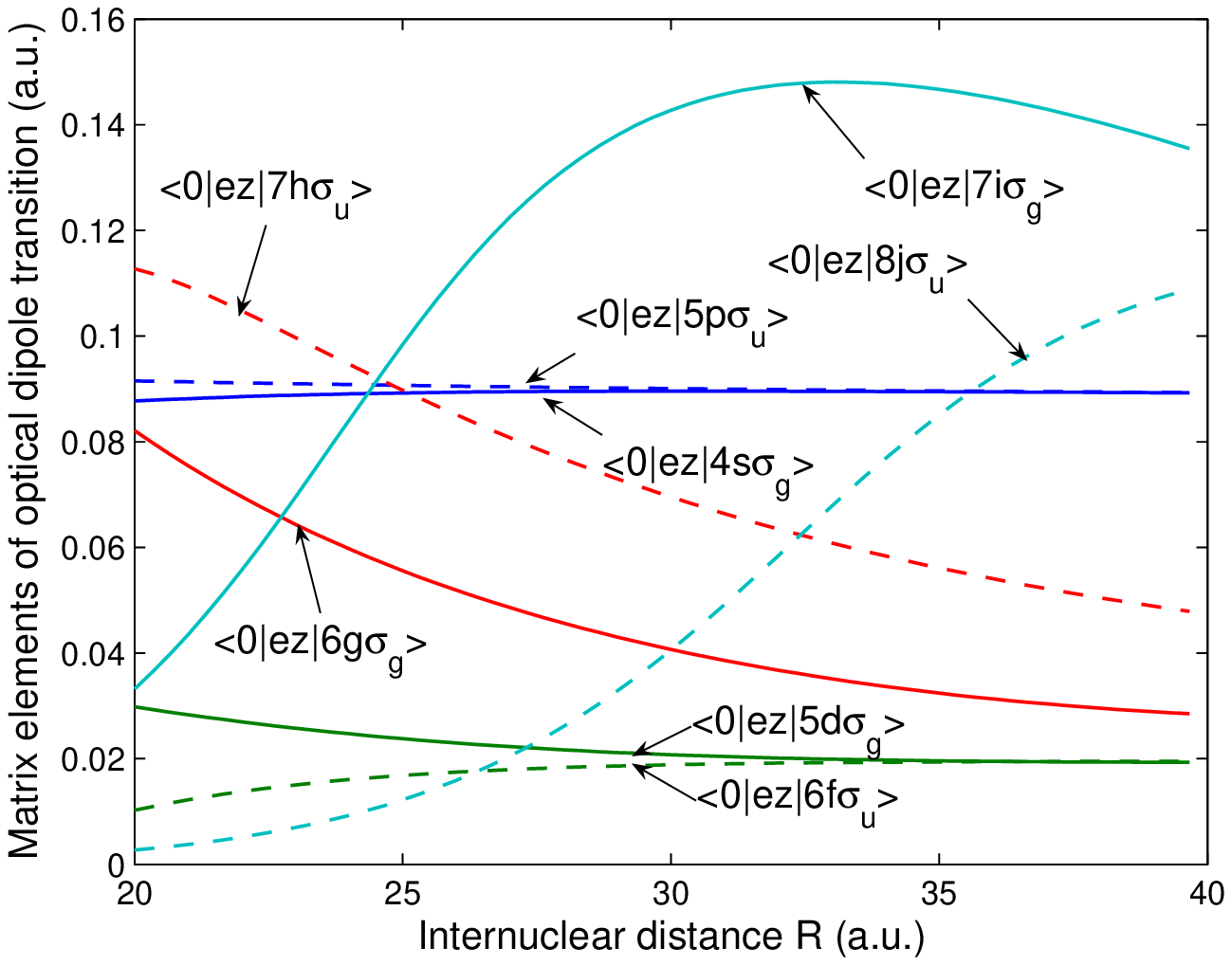}}
 \centerline{Fig. 4 (c)}

\newpage
\centerline{\includegraphics[width=12cm]{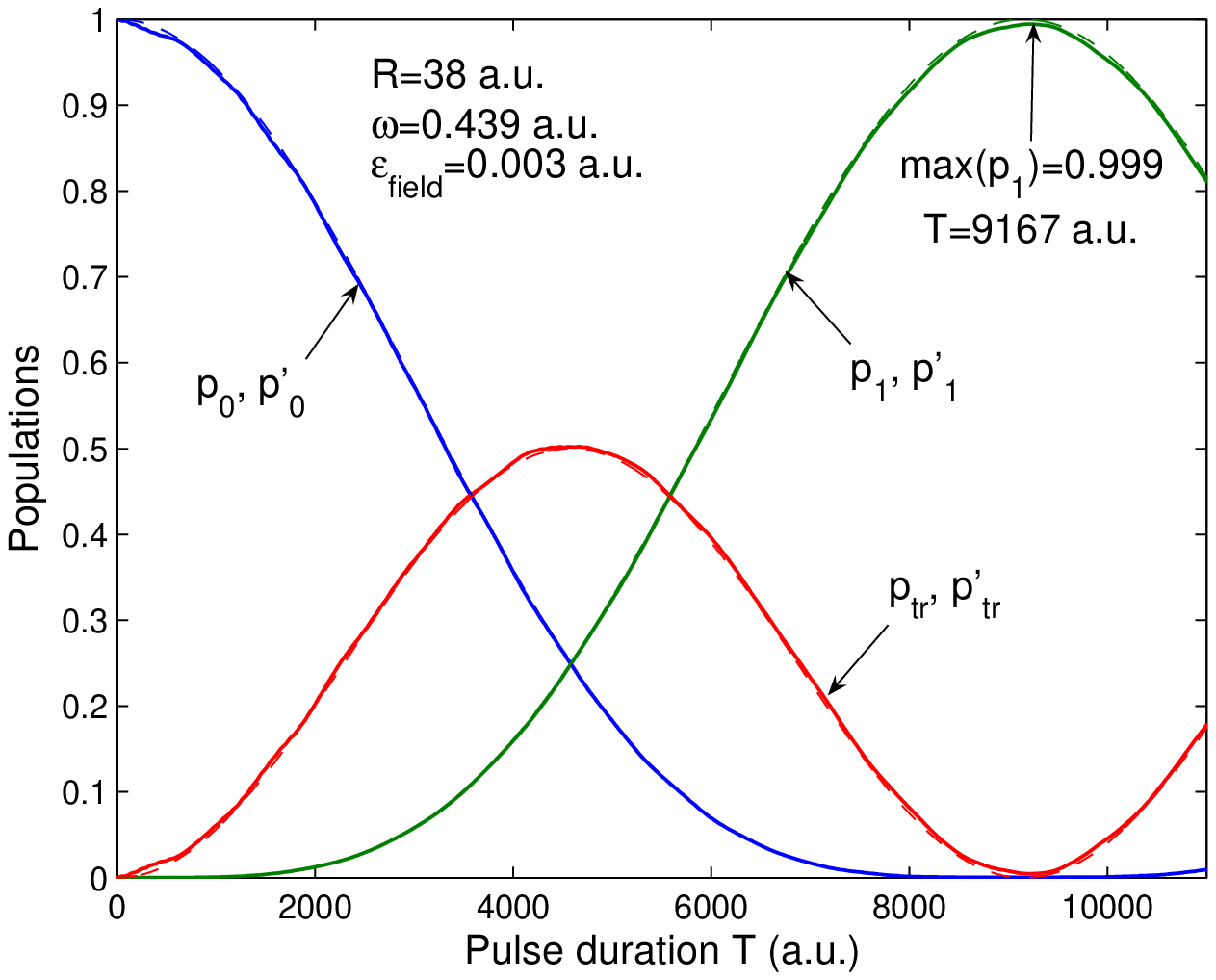}}
 \centerline{Fig. 5}

\centerline{\includegraphics[width=12cm]{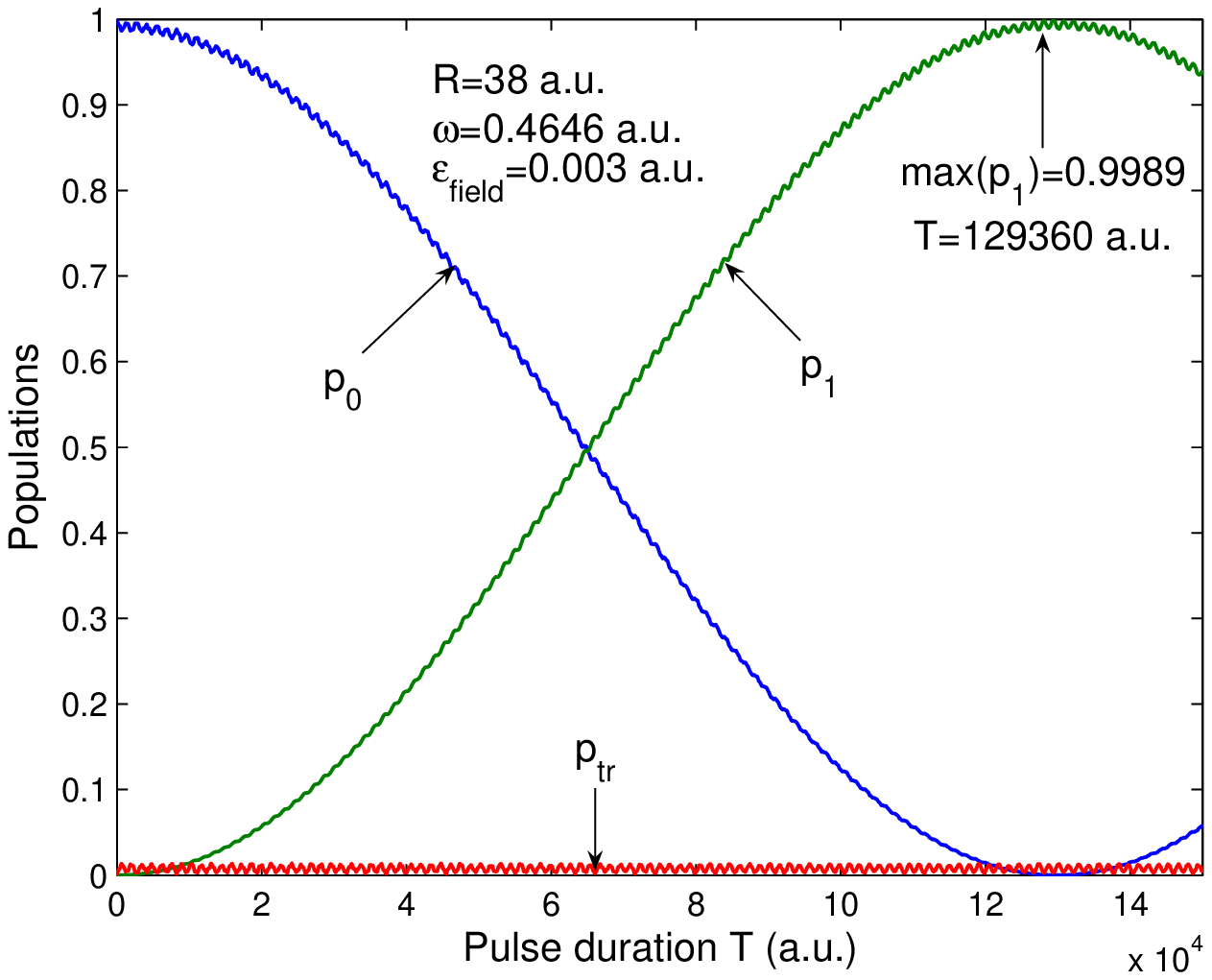}}
 \centerline{Fig. 6}

\newpage
\centerline{\includegraphics[width=12cm]{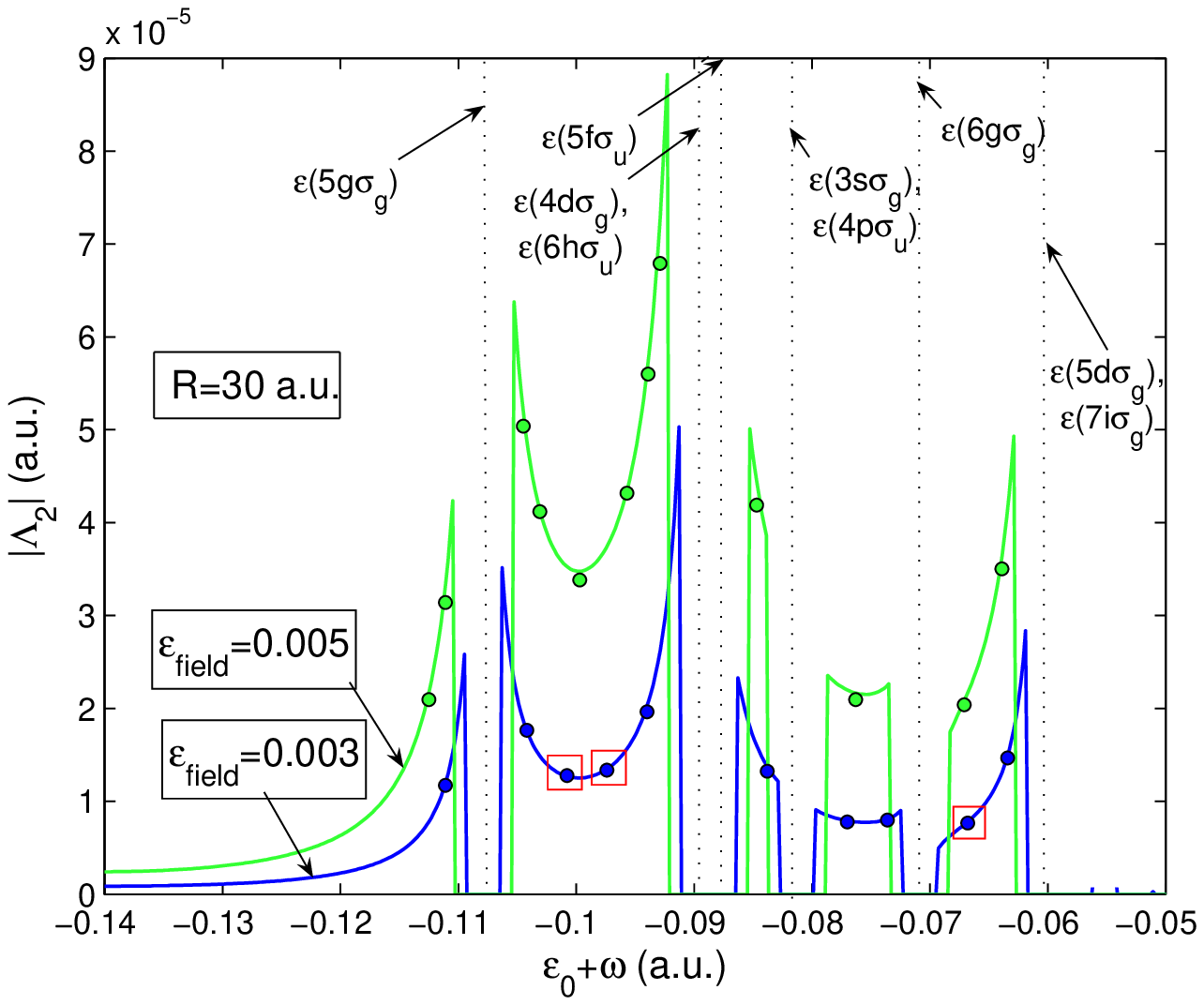}}
 \centerline{Fig. 7 (a)}

\centerline{\includegraphics[width=12cm]{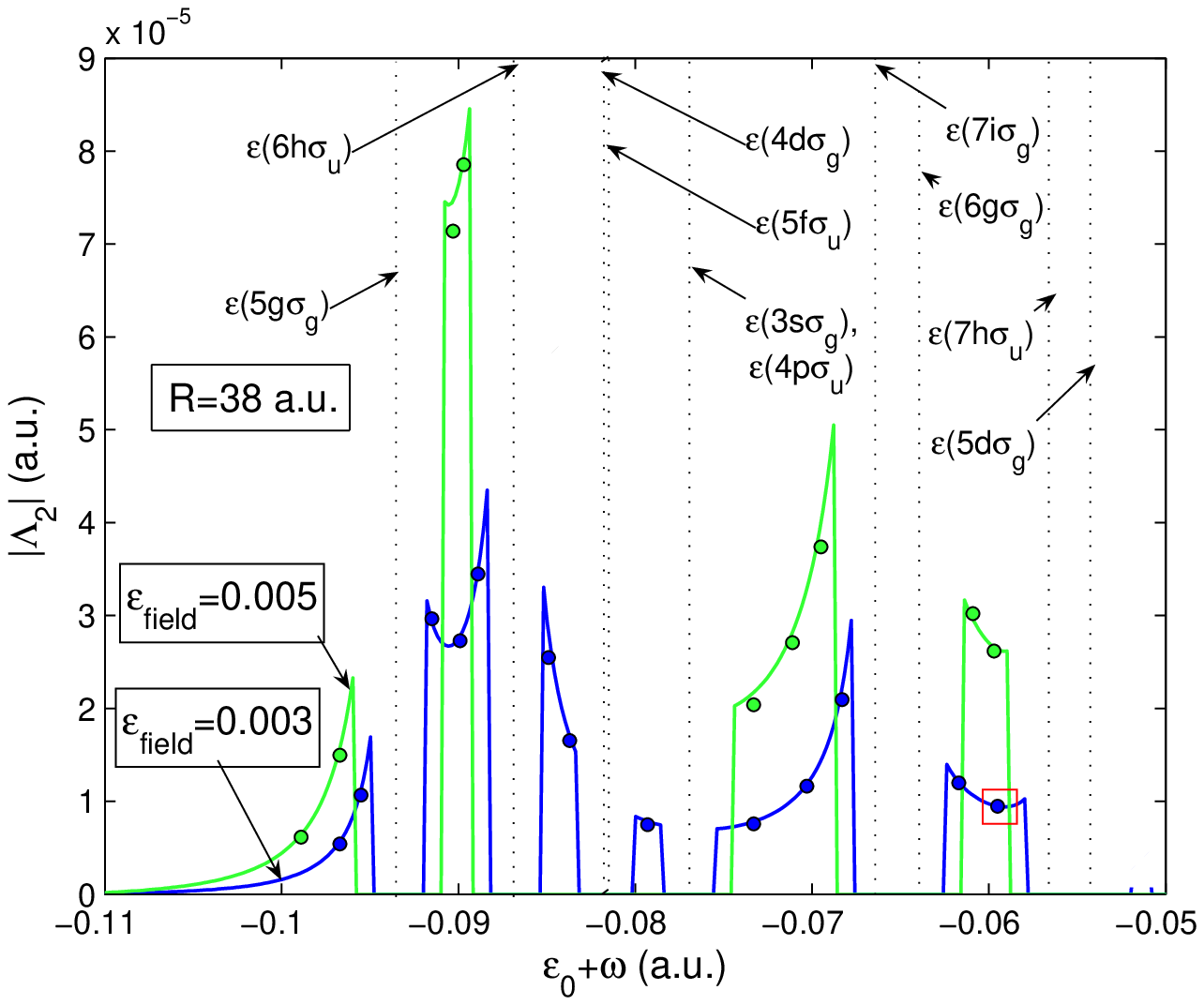}}
 \centerline{Fig. 7(b)}

\newpage
\centerline{\includegraphics[width=12cm]{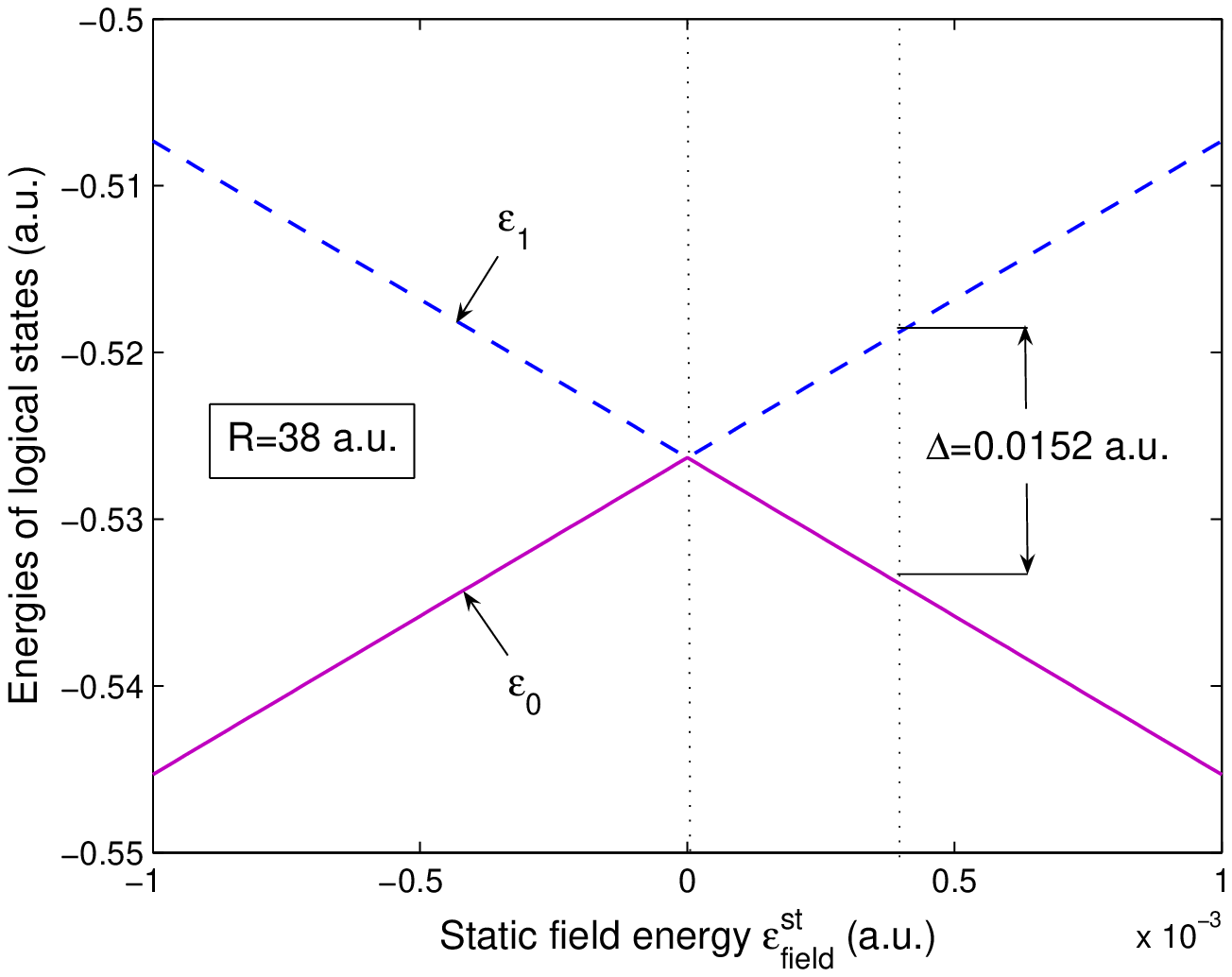}}
 \centerline{Fig. 8 (a)}

\centerline{\includegraphics[width=12cm]{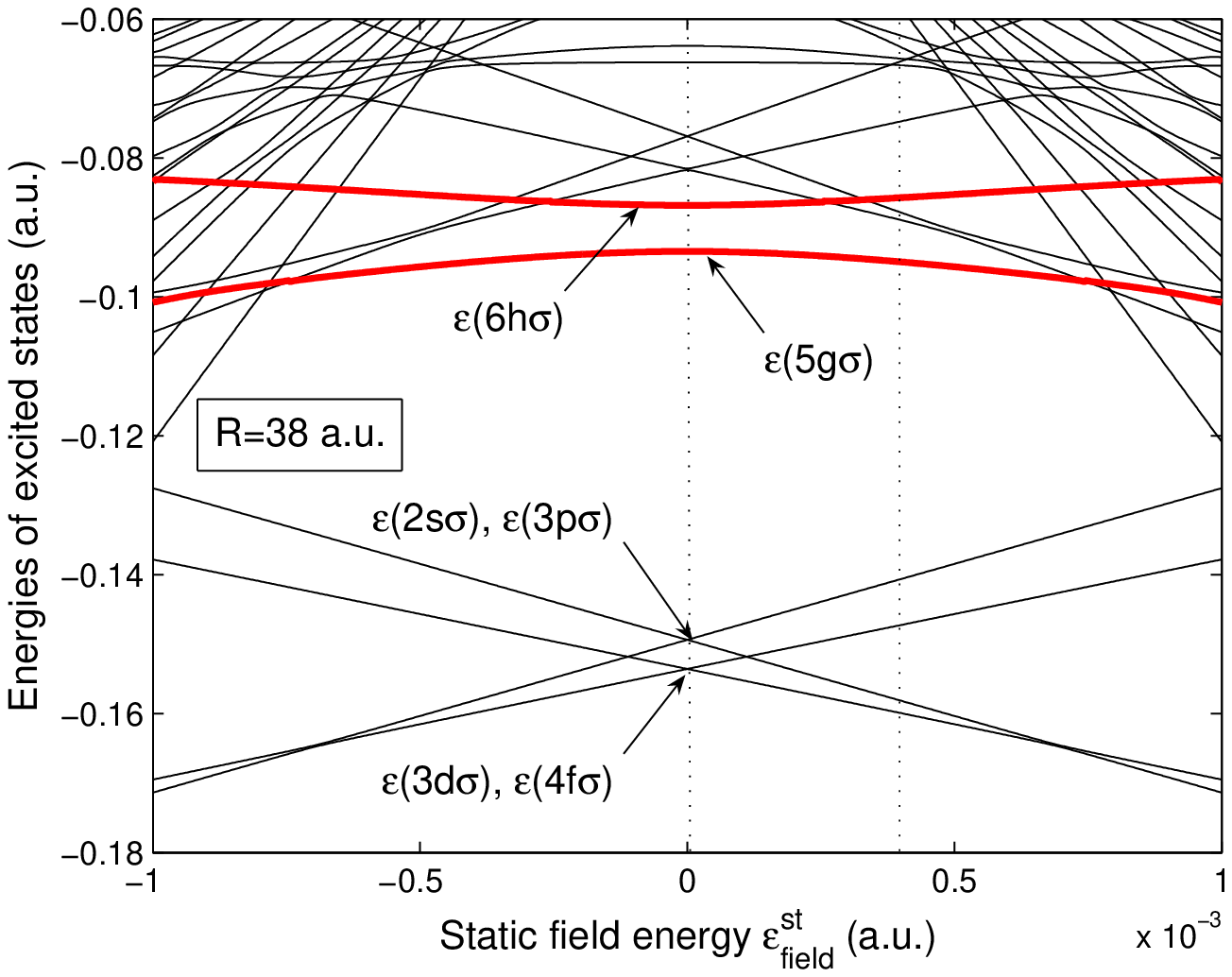}}
 \centerline{Fig. 8 (b)}

\newpage
\centerline{\includegraphics[width=12cm]{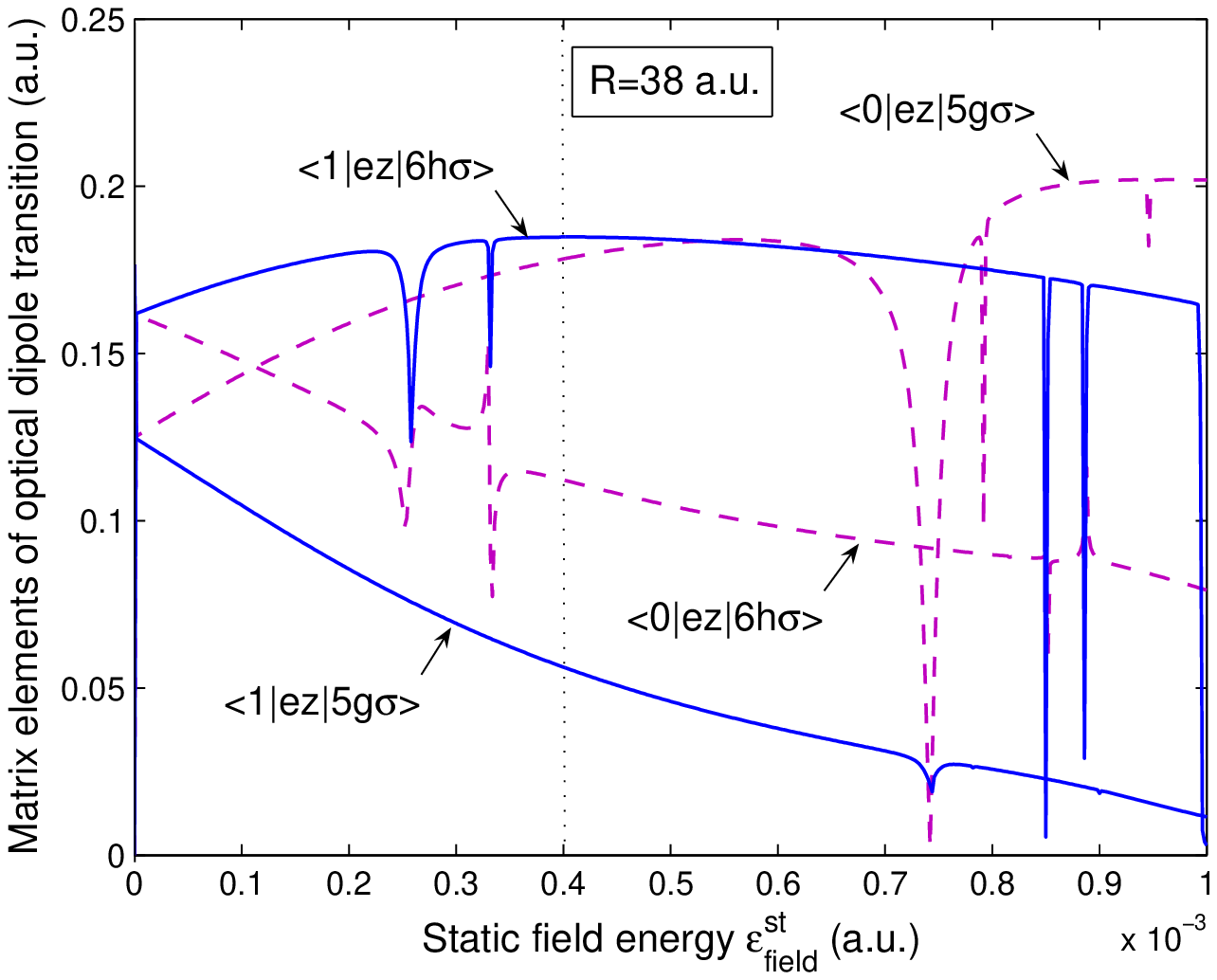}}
 \centerline{Fig. 9}

\centerline{\includegraphics[width=12cm]{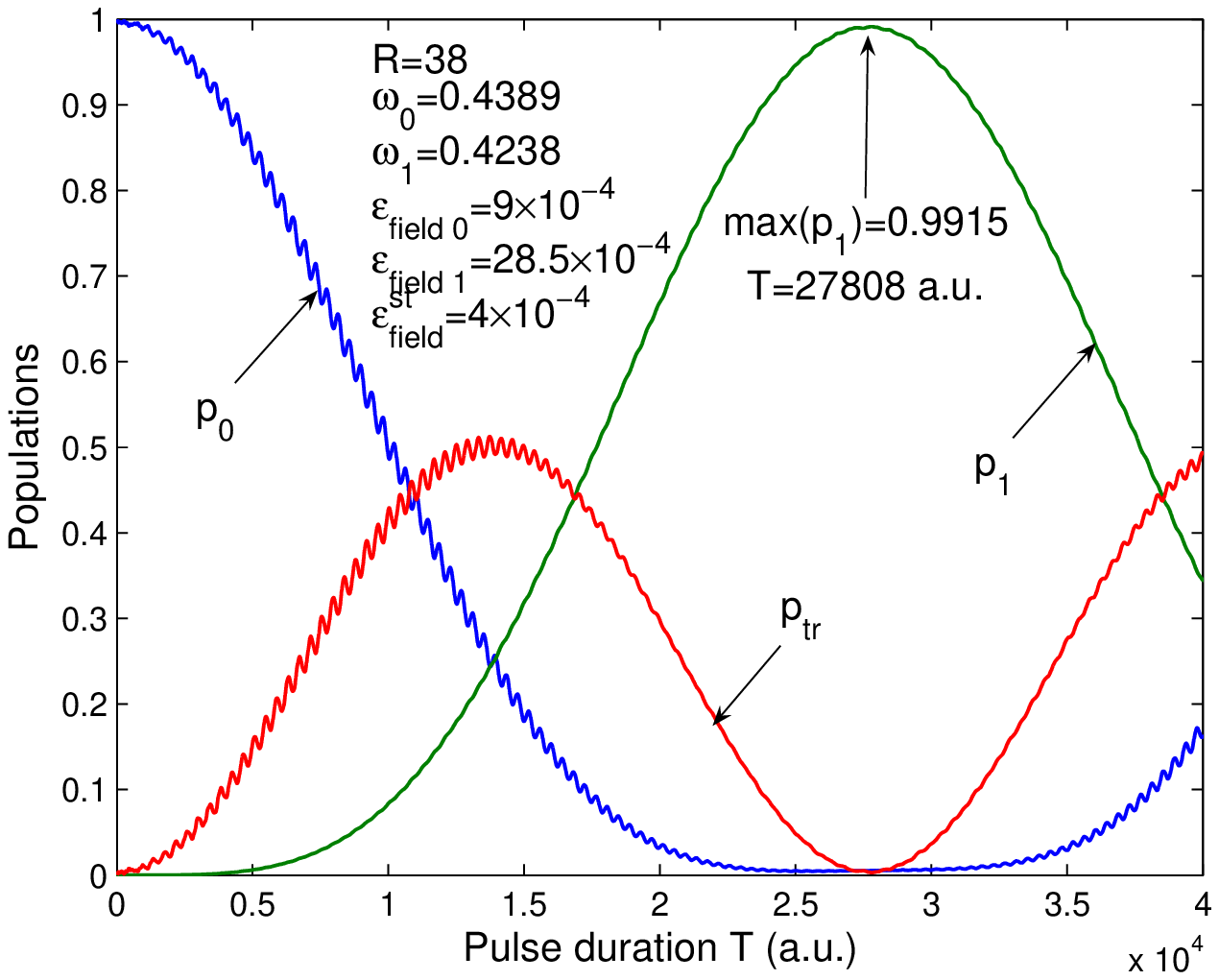}}
 \centerline{Fig. 10}

\newpage
\centerline{\includegraphics[width=12cm]{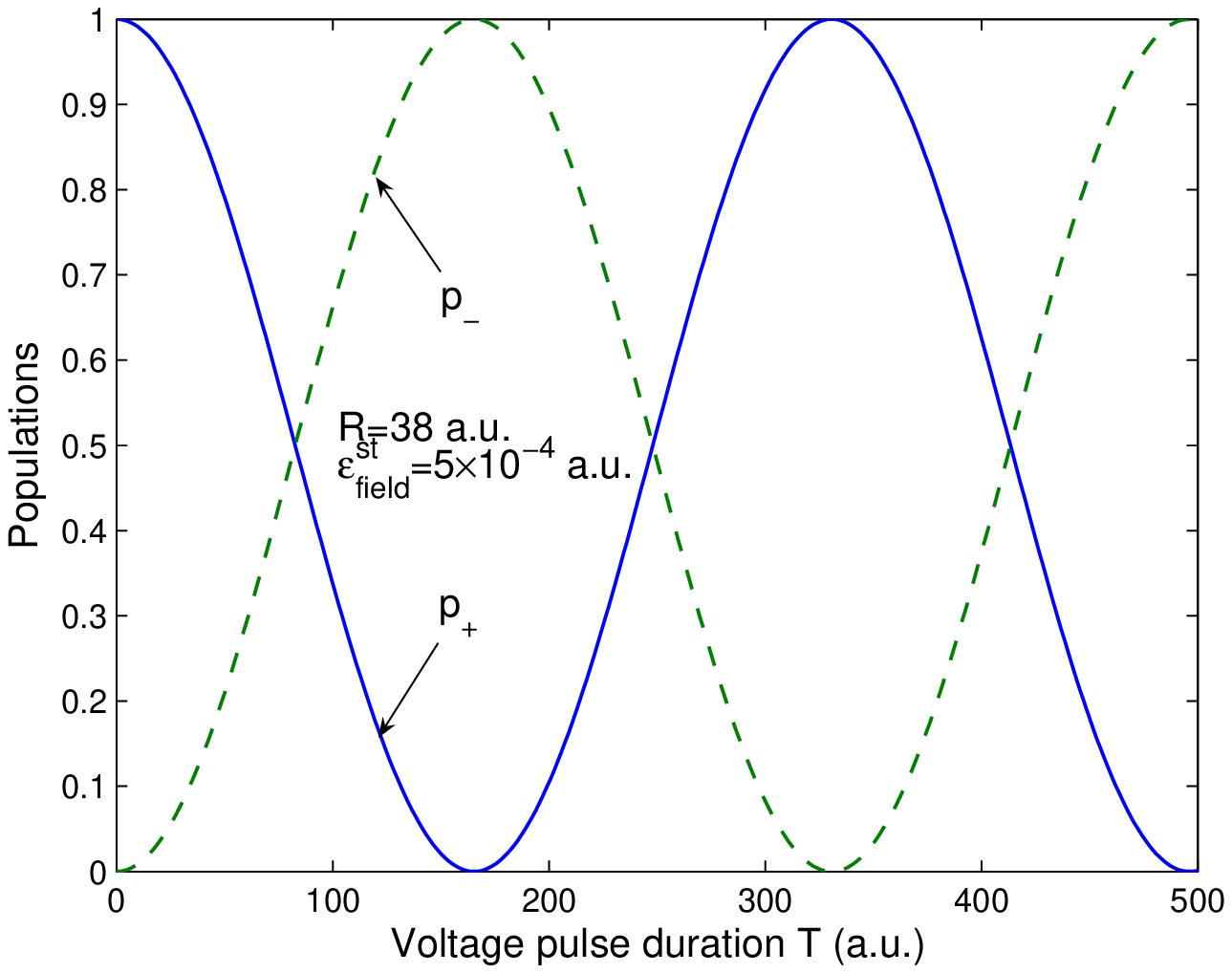}}
 \centerline{Fig. 11}

\end{document}